\documentclass{aa}
\usepackage{graphicx}
\usepackage{txfonts}
\newcommand{\kpc}{\rm{\, kpc}}
\newcommand{\kms}{\rm{\, km\, s^{-1}}}
\newcommand{\Kkms}{\rm{\, K\, km\, s^{-1}}}
\newcommand{\pheins}{\phantom{1}}
\begin{document}
\title{Molecular gas in the Andromeda galaxy}

\author{Ch.~Nieten\inst{1}\fnmsep\thanks{\emph{Present address:} c.nieten@zeiss.de}
        \and N.~Neininger\inst{1,2,3}\fnmsep\thanks{\emph{Present address:} NNeininger@kpmg.com}
        \and M.~Gu\'elin\inst{3} \and H.~Ungerechts\inst{4}
        \and R.~Lucas\inst{3}
        \and E.~M.~Berkhuijsen\inst{1}
        \and R.~Beck\inst{1}
        \and R.~Wielebinski\inst{1} }

\offprints{guelin@iram.fr}

\authorrunning{Ch.~Nieten et al.}
\titlerunning{Molecular gas in the Andromeda galaxy}

\institute{Max-Planck-Institut f\"ur Radioastronomie,
           Auf dem H\"ugel 69,
           53121 Bonn, Germany
        \and Radioastronomisches Institut der Universit\"at Bonn,
           Auf dem H\"ugel 71,
           53121 Bonn, Germany
        \and Institut de Radioastronomie Millim\'etrique,
           300, rue de la piscine, 38406 St. Martin d' H\`eres,
           France
        \and Instituto de Radioastronom\'{\i}a Milim\'etrica,
           Avenida Divina Pastora 7, 18012 Granada, Spain
}

\date{Received 13 November 2003 / Accepted 5 December 2005}

\abstract{
We present a new \element[][12]{CO}(J=1--0)--line survey of the
Andromeda galaxy, M\,31, with the highest resolution to date
($23\arcsec$, or
85~pc along the major axis), observed {\em On-the-Fly} with the IRAM
30-m telescope. We mapped an area of about $2\degr\times 0\fdg 5$ which
was tightly sampled on a grid of $9\arcsec$ with a velocity resolution
of $2.6\kms$. The r.m.s. noise in the velocity-integrated map is around
$0.35\Kkms$ on the $T_\mathrm{mb}$-scale.\\
Emission from the \element[][12]{CO}(1--0) line is detected from
galactocentric radius $R=3$\,kpc to $R=16$\,kpc, but peaks in intensity
at $R\sim 10$\,kpc. Some clouds are visible beyond $R=16$\,kpc, the
farthest of them at $R=19.4$\,kpc.  \\
The molecular gas traced by the (1--0) line is concentrated in narrow
arm-like filaments, which often coincide with the dark dust lanes
visible at optical wavelengths. The \ion{H}{i} arms are broader and
smoother than the molecular arms. Between $R=4$\,kpc and $R=12$\,kpc
the brightest CO filaments and the darkest dust lanes define a
two-armed spiral pattern that is well described by two logarithmic
spirals with a constant pitch angle of 7\degr--8\degr. Except for some
bridge-like structures between the arms, the interarm regions and the
central bulge are free of emission at our sensitivity. The
arm--interarm brightness ratio averaged over a length of 15~kpc along
the western arms reaches about 20 compared to 4 for \ion{H}{i} at an
angular resolution of $45\arcsec$.\\
In several selected regions we also observed the
\element[][12]{CO}(2--1)--line on a finer grid. Towards the bright CO
emission in our survey we find normal ratios of the (2--1)--to--(1--0)
line intensities which are consistent with optically thick lines and
thermal excitation of CO.   \\
We compare the (velocity-integrated) intensity distribution of CO with
those of \ion{H}{i}, FIR at $175\,\mu$m and radio continuum,
and interpret the CO data in terms of molecular gas column densities.
For a constant conversion factor $X_\mathrm{CO}$, the molecular
fraction of the neutral gas is enhanced in the spiral arms and
decreases radially from 0.6 on the inner arms to 0.3 on the arms at
$R\simeq 10$~kpc. We also compare the distributions of \ion{H}{i},
H$_2$ and total gas with that of the cold (16\,K) dust traced at
$\lambda175\,\mu$m. The ratios $N(\ion{H}{i})/I_{175}$ and
$(N(\ion{H}{i})+2N(\mathrm{H}_2))/I_{175}$ increase by a factor of
$\sim20$ between the centre and $R\simeq 14\kpc$, whereas the ratio
$2N(\mathrm{H}_2)/I_{175}$ only increases by a factor of 4. For a
constant value of $X_\mathrm{CO}$, this means that either the atomic and
total gas--to--dust ratios increase by a factor of $\sim20$ or that the
dust becomes colder towards larger radii. A strong variation of
$X_\mathrm{CO}$ with radius seems unlikely. The observed
gradients affect the cross-correlations between gas
and dust. In the radial range $R=8$--14~kpc total gas and cold dust are
well correlated; molecular gas is better correlated with cold dust than
atomic gas. At smaller radii no significant correlations between gas
and dust are found.\\
The mass of the molecular gas in M\,31 within a radius of 18~kpc is
$M (\mbox{H}_2) = 3.6\times 10^8\, \mbox{M}_{\sun}$ at the adopted
distance of 780~kpc. This is 12\% of the total neutral gas mass within
this radius and 7\% of the total neutral gas mass in M\,31.
}

\maketitle
%
%________________________________________________________________

\section {Introduction}

Star formation and spiral structure in galaxies require the coupling of
processes operating on linear scales so different that they are hard to
study in a single galaxy. The small structures are difficult to observe
in external galaxies, whereas large structures, due to distance
ambiguities, are hard to see in the Milky Way. Single-dish telescopes
were used to survey CO in galaxies (e.g. Nakano et al.\ \cite{nakano+87};
Braine et al.\ \cite{braine+93}; Young et al.\ \cite{young+95})
but with limited angular resolution.
Molecular spiral arms were barely resolved in these surveys even in the
nearest galaxies (e.g. Koper et al.\ \cite{koper+91}; Garcia-Burillo et
al.\ \cite{garcia-burillo+93}; Loinard et al.\ \cite{loinard+96}; Heyer
et al.\ \cite{heyer+04}). Only in the Magellanic Clouds the single-dish
surveys resolved giant molecular clouds (Israel et al.\
\cite{israel+93}). More recently
mm-wave interferometer surveys like the BIMA SONG (Regan et al.\
\cite{regan+01}) gave vastly improved data on nearby galaxies like
M\,51, resolving molecular arms into cloud complexes. This
instrument was also used for an all-disk survey of M\,33, about ten
times closer to us than M\,51, in which individual molecular
clouds are recognized (Engargiola et al.\ \cite{engarg+03}).
The IRAM Plateau de Bure interferometer has resolved molecular clouds
in M\,31 into components (Neininger et al.\ \cite{neininger+00a})
enabling close comparisons with molecular clouds in the Milky Way.

The nearest large spiral is the Andromeda Nebula, M\,31. Its distance
of $D= 0.78\pm 0.04$~Mpc (Stanek \& Garnavich\
\cite{stanek+garnavich98}) ranks among the best known for any galactic or
extragalactic nebula. The accuracy of this distance allows us to derive
accurate luminosities and masses. At this distance $1\arcmin$ along the
major axis corresponds to $227\pm 12$~pc. The large inclination of
M\,31, $i= 77\fdg 5$, degrades the resolution along the minor axis by a
factor of 4.6, but has the advantage of yielding accurate in-plane
velocities. M\,31's proximity gives us the chance to see many details
of the  distribution and kinematics of the gas as well as the relation
of the gas to the spiral structure and to star formation.

The contents of stars, dust, and atomic gas in M\,31 are well known.
The whole galaxy has been mapped in the 21~cm line of \ion{H}{i} with
$24\arcsec\times 36\arcsec$ resolution by Brinks \& Shane
(\cite{brinks+shane84}, hereafter B\&S) and its northeastern half with
$10\arcsec$ resolution by Braun (\cite{braun90}). It has been
entirely mapped in the mid and far infrared by the IRAS, ISO and Spitzer
satellites (see Haas et al. (\cite{haas+98}) and Schmidtobreick et al.
(\cite{schmidto+00}) for the ISOPHOT map at $175\,\mu$m,
and Gordon et al. (\cite{gordon+04}) for the MIPS maps at
$20\,\mu$m, $60\,\mu$m and $160\,\mu$m). Furthermore, M\,31 was
partially mapped with ISOCAM (5.1--$16.5\,\mu$m) at $6\arcsec$
resolution (see e.g. Pagani et al.\ \cite{pagani+99}). Comparisons of
the emission in different wavelength ranges -- like UV, optical,
\ion{H}{i}, FIR ($160\,\mu$m and $175\,\mu$m) and radio continuum
emissions --  have also been reported (Loinard et al.\
\cite{loinard+99}; Pagani et al.\ \cite{pagani+99};
Keel\ \cite{keel00}; Lequeux\ \cite{lequeux00}; Nieten et al.\
\cite{nieten+00}; Berkhuijsen et al.\ \cite{elly+00}; Gordon et al.\
\cite{gordon+04}).

So far, the situation was not as favourable as to the molecular gas. Prior
to ours, the only complete CO survey of M\,31 was made with a 1.2-m
diameter telescope and had a resolution of $8\farcm 7$ (Koper et
al.\ \cite{koper+91}; Dame et al.\ \cite{dame+93}). More recently, a
survey of the southwestern half, made at an angular resolution of
$1\arcmin$ with the FCRAO 14-m telescope, was published by Loinard et
al. (\cite{loinard+96}, \cite{loinard+99}). The latter authors
(\cite{loinard+99}, their Table~2) give a nearly complete overview of
previous CO observations of M\,31. Loinard et al. (\cite{loinard+99})
and Heyer et al. (\cite{heyer+00}) found many similarities, but also
clear differences, between properties of the molecular gas in M\,31
and those in the Milky Way.

Our survey, made with the IRAM 30-m telescope in the
\element[][12]{CO}(J=1--0) line, has a resolution of $23\arcsec$
corresponding to 85~pc along the major axis. It is much more sensitive
than the previous surveys and detects all clouds with
$\int T_\mathrm{mb} {\rm dv} \ga 1\Kkms$ (= $3\times$ rms noise).
In this article we present the CO distribution
in the bright disk of the galaxy. We derive some important basic
results using simple assumptions, e.g., a constant conversion factor
$X_\mathrm{CO}$ from CO intensity to molecular column density.
We discuss the spiral-arm structure of the neutral gas
and the arm-interam brightness contrast in Sect.~3. In addition
to the \element[][12]{CO}(J=1--0) line, we have observed several
selected areas covering bright arm segments, in the
\element[][12]{CO}(J=2--1) line with high sensitivity; we discuss the
line ratios in Sect.~3.3. In a previous publication, based on one third
of the present data (Neininger et al.\ \cite{neininger+98}), we reported a
tight correlation between the CO sources and the dark dust lanes.
In Sect.~4 we return to this point and compare the CO distribution with
those of \ion{H}{i}, FIR ($175\,\mu$m) and $\lambda$20~cm radio
continuum. Radial profiles of the various constituents are discussed in
Sect.~4.1 and correlations between CO, \ion{H}{i} and FIR ($175\,\mu$m)
in Sect.~4.2. In Sect.~4.3 we derive the molecular and total gas mass.
The CO velocity field is described in Sect.~5. Our results are
summarized in Sect.~6. Preliminary reports on this survey
were given by Gu\'elin et al. (\cite{guelin+00}), Neininger et al.
(\cite{neininger+98}, \cite{neininger+00b}) and Nieten et al.
(\cite{nieten+00}).

%--------------------------
\section {Observations}

Our survey was carried out with the IRAM 30-m telescope between
November 1995 and August 2001. The observations were made {\em
On-the-Fly} in two steps: in a first step a field typically
$18\arcmin\times 18\arcmin$ in size was scanned back and forth in the
direction parallel to M\,31's minor axis, $Y$, at a speed of $\rm
4\arcsec\,s^{-1}$. The successive scans were spaced by $9\arcsec$ in
the orthogonal $X$ direction. At the beginning and at the end of each
scan, a reference position, located $30\arcmin$ or $45\arcmin$ away
from the major axis\footnote{Throughout this paper we use the
M\,31-fixed coordinate system of Baade \& Arp
(\cite{baade+arp64}): centre coordinates $\rm RA = 0^h40^m 00\fs 3,\
DEC = 41\degr 00\arcmin 03\arcsec$ (1950.0) (Dressel \& Condon\
\cite{dressel+condon76}), $X$ along the major axis ($PA= 37\fdg 7$),
positive to the NE, $Y$ along the minor axis, positive to the SE. All
velocities are in the LSR frame.} and free of CO or \ion{H}{i} emission,
was observed for 30~s. Every 1--2~hours the telescope pointing was
checked on planets and nearby quasars. A second reference position,
located within M\,31 and showing strong CO emission, was observed for
calibration purposes (see below).  The telescope focus was checked
several times a day, in particular
after sunrise and sunset. In a second step the observations were
repeated by scanning the same field in the orthogonal direction,
parallel to the major axis. The data recorded by the backends were read
every second of time, so that the data cube obtained by combining the
two orthogonal maps was fully sampled on a $9\arcsec$ grid.

The reduction procedure was described in some detail by Neininger
et al. (\cite{neininger+00b}). After calibration (see below),
subtraction of the off-source reference spectrum and a baseline
for individual spectra in a map, two orthogonal maps were combined
using ``basket-weaving'', the de-striping technique of Emerson \&
Gr\"ave (\cite{emerson+graeve88}). This code was adapted to
work on two orthogonal channel maps before averaging them
(Hoernes\ \cite{hoernes97}). Examples of the technique are shown in
Neininger et al. (\cite{neininger+00b}), Hoernes (\cite{hoernes97}) and
Emerson \& Gr\"ave (\cite{emerson+graeve88}). In  this process, the CO
map was smoothed from $21\arcsec$ to $23\arcsec$ FWHM.

\begin{figure*}[p]
\begin{center}
\includegraphics[bb = 98 93 528 530,width=18cm,clip=]{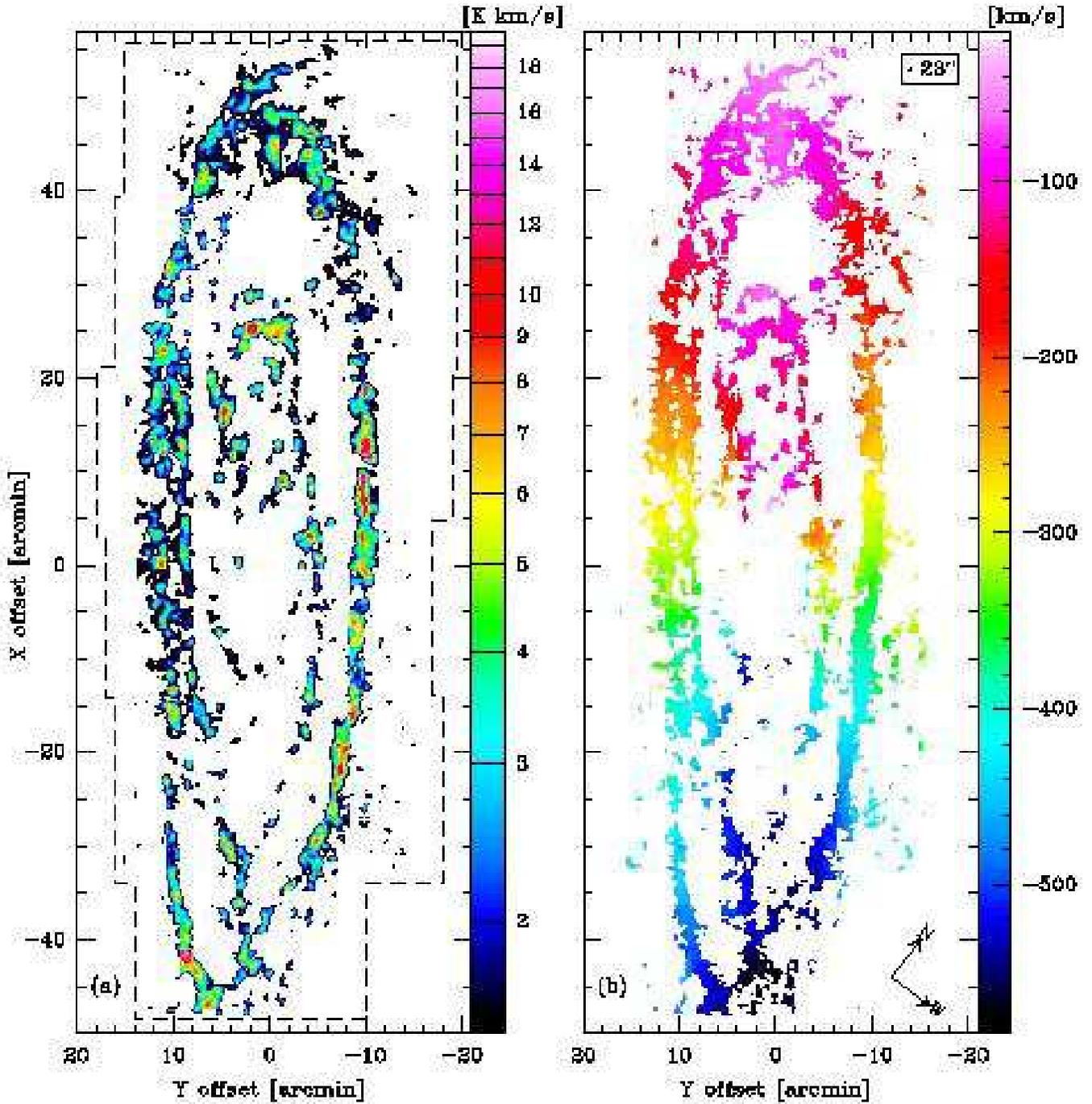}
\caption{ \label{fig:co-map}
{\bf (a)} The velocity-integrated intensity distribution of the
\element[][12]{CO}(1--0) spectrum, $I_{1-0}=\int T_\mathrm{mb}
(CO_{1-0})\mathrm{dv}$, observed with the IRAM 30-m telescope. The $X$
and $Y$ coordinates are taken along the major and minor axis,
respectively; the position angle and centre position are taken from
Dressel \& Condon (\cite{dressel+condon76} -- see footnote 1). The
dashed line marks the border of the area surveyed which is about one
degree squared. The colour scale starts at $1.5\Kkms$. Most structures
visible in the map, including weak ones, correspond to line profiles
detected with signal-to-noise ratios $\geq 4$. The mean r.m.s. noise in
the integrated intensities is $0.35\Kkms$, but the noise varies across
the map (see Table~\ref{tab:data}).
{\bf (b)} The velocity field (first moment) as traced by the CO
emission. Velocities are LSR-values.}
\end{center}
\end{figure*}

The maps shown in Fig.~\ref{fig:co-map} are the combination of
12 individual fields, listed in Table~\ref{tab:data}. Each field is
fully sampled and all fields together contain nearly 1.7 million spectra
(before gridding) obtained in about 500 hours of effective observing
time. We present individual spectra for selected regions in
Fig.~\ref{fig:spectra}.

%Fig. 1

\clearpage

When observing the CO(1--0) line, the 30-m telescope allows
simultaneous observations of two polarizations of the (2--1) line.
The (2--1) line, however, is weaker than the (1--0) line and the
integration time per beam is about $4\times$ smaller because the beam
area is $4\times$ smaller. Moreover, the receivers are
noisier and the sky opacity is higher at 230~GHz (2--1) than at 115~GHz
(1--0). This made it difficult to detect the (2--1)-line emission in
our survey, except from the brightest clouds. In order to improve the
signal-to-noise ratio and the sampling for (2--1) we re-observed several
rectangular regions, $3\arcmin$--$4\arcmin$ wide by
$3\arcmin$--$12\arcmin$ long, with half the scanning velocity and twice
the sampling, when the zenith opacity at 230~GHz was favourably low
($\leq 0.2$). We discuss some general results of these observations in
Sect.~3.3.
In order to confirm the reliability of the OTF method, to integrate some
emission-free positions to a lower noise level, and to check several
apparent discrepancies with previous CO observations we re-observed
about 200 positions, located inside as well as outside the arms,
in the position-switching mode with integration times of
10--30~min. Some of these results are used in Sect.~3.2.\footnote{The
full results of the position-switched observations will be presented
elsewhere.}

%Table 1
\begin{table*}[htb]
\begin{center}
\caption{Data on observed fields of M\,31}
\label{tab:data}
\begin{tabular}{rcr@{ ... }rr@{ ... }rrcc}
\hline\hline
\noalign{\smallskip}
Field  &Quadrant  &\multicolumn{2}{c}{$X$-extent}
       &\multicolumn{2}{c}{$Y$-extent} &Number
       &\multicolumn{2}{c}{Mean r.m.s. noise}\\
\#     &in M\,31 &\multicolumn{2}{c}{} &\multicolumn{2}{c}{}
       &of spectra &in $I_{1-0}$  &per channel\\
       &  &\multicolumn{2}{c}{(\arcmin)} &\multicolumn{2}{c}{(\arcmin)}
       & &$\rm (K\, km\, s^{-1})$ &(mK)$^{1)}$\\
\noalign{\smallskip}
\hline\noalign{\smallskip}
 1  &S  &$-$48.5&$-$33.3  &$-$10.1&14.0     &162980  &0.45 &34\\
 2  &SW &$-$34.0&$-$13.9  &$-$18.1&$-$2.0   &225122  &0.38 &37\\
 3  &SE &$-$34.0&$-$13.9  &2.0&16.1         &115765  &0.35 &34\\
 4  &SW &$-$14.2&5.1      &$-$17.0&$-$3.0   &128027  &0.44 &33\\
 5  &NE &21.0&39.3        &$-$2.0&16.1      &204190  &0.46 &32\\
 6  &NW &21.0&39.1        &$-$19.6&$-$1.5   &152630  &0.25 &22\\
 7  &NE &39.0&55.9        &$-$2.0&15.1      &160045  &0.27 &21\\
 8  &NW &38.9&56.1        &$-$19.6&$-$1.5   &117198  &0.29 &27\\
 9  &NE &3.0&21.1         &$-$2.0&18.0      &164760  &0.32 &25\\
10  &SE &$-$14.1&3.2      &$-$3.0&17.0      &128534  &0.27 &27\\
11  &NW &4.8&21.1         &$-$19.0&$-$1.0   &128156  &0.33 &24\\
12  &NW &2.8&5.1          &$-$3.0&$-$1.8    &666     &0.36 &26\\
\noalign{\medskip}
    &NE &\multicolumn{2}{c}{$X>0$} &\multicolumn{2}{c}{$Y>0$}
        &528995 &\hspace{0.5em} 0.36$^{2)}$ &\quad 26$^{2)}$\\
    &NW &\multicolumn{2}{c}{$X>0$} &\multicolumn{2}{c}{$Y<0$}
        &398650  &0.29  &24\\
    &SE &\multicolumn{2}{c}{$X<0$} &\multicolumn{2}{c}{$Y>0$}
        &339371  &0.35  &31\\
    &SW &\multicolumn{2}{c}{$X<0$} &\multicolumn{2}{c}{$Y<0$}
        &421057  &0.41  &35\\
\noalign{\smallskip}
\hline
\noalign{\smallskip}
 &Total &\multicolumn{4}{l}{3512 square arcminutes}  &1688073
      &0.35 &29\\
\noalign{\smallskip}
\hline
\noalign{\medskip}
\multicolumn{9}{l}{$^{1)}$ Temperatures are on $T_\mathrm{mb}$-scale,
   where $T_\mathrm{mb} = 1.15 T_\mathrm{A}^*$.
   $^{2)}$ Weighted means}\\
\end{tabular}
\end{center}
\end{table*}

We used two SIS receivers with orthogonal polarizations to observe the
(1--0) line and a similar system to observe the (2--1) line. The
receiver temperatures in the standard reference plane (before
the polarization splitter) were close to 90~K (SSB) at 115~GHz at the
beginning of our survey (fields 1--3, see Table~\ref{tab:data}) and
close to 50~K at the end (fields 8--11). After addition of the
atmospheric contribution, the system temperature was between 200 and
400~K at both 115~GHz and at 230~GHz. The backends consisted of two
$512\times 1$~MHz filterbanks at 115~GHz and of 2 autocorrelators with
resolutions of 0.8~MHz and total bandwidth of 320~MHz at 230~GHz.
A channel width of 1~MHz corresponds to a velocity resolution of
$2.6\kms$ for the (1--0) line and of $1.3\kms$ for the (2--1) line.

The standard calibration at the 30-m telescope is equivalent to the
chopper-wheel calibration method for observations at millimeter
wavelengths and gives the antenna temperature $T_\mathrm{A}^*$,
corrected for atmospheric losses and forward efficiency,
$F_\mathrm{eff}$ (for details see Downes\ \cite{downes89}). The
main-beam brightness temperature, $T_\mathrm{mb}$, can then be
calculated from
\begin{equation}
T_\mathrm{mb} = F_\mathrm{eff} / B_\mathrm{eff} T_\mathrm{A}^*\ ,
\end{equation}
where $B_\mathrm{eff}$ is the main-beam efficiency.
Over the years of this project, $F_\mathrm{eff}$ and in particular
$B_\mathrm{eff}$ improved significantly due to upgrades of the 30-m
telescope, the most important of which occurred in July 1997. We used
the standard values for these efficiencies for the last quarter of 1997
derived by the observatory staff, for $F_\mathrm{eff}$ from antenna
tippings and for $B_\mathrm{eff}$ from observations of Mars and Uranus:
$F_\mathrm{eff}/B_\mathrm{eff} = 0.92/0.80$ at 115~GHz
and $0.80/0.50$ at 230~GHz.
Day--to--day variations were monitored by observing two reference
positions: $X,Y = -30\farcm 18,\ 4\farcm 26$ in the south and $X,Y =
18\farcm 80,\ 1\farcm 90$ in the north. Normalization of the line
integrals at these positions to the same values ensured the same
calibration parameters for the entire survey. The uncertainty in the
final $T_\mathrm{mb}$ scales is about 15\%.

A corollary of the high beam efficiency of the 30-m telescope at
115~GHz is the low error beam. Greve et al. (\cite{greve+98}) found that
the far-beam pattern can be described by 3 error beams with half-power
widths of $5\arcmin$, $7\arcmin$ and $80\arcmin$, respectively. The
first two could pick up signals from regions with similar radial
velocities as observed in the main beam. As they contribute only a few
percent to the main-beam signals, they hardly affect the observed
spectra. The contribution of the third error beam, which is larger than
the disk of M\,31, is negligible.

The effective integration time per $23\arcsec$ beam for the large
\element[][12]{CO}(1--0) map was 64~s,
yielding a r.m.s. noise of $\simeq\, 33$~mK per 1~MHz channel
($T_\mathrm{mb}$ scale) in the southern fields (1--4) and $\simeq\,
25$~mK per 1~MHz in the northern ones (see Table~\ref{tab:data}). The
corresponding values for the small maps are $21\arcsec$, 173~s
and 15~mK for the (1--0) line, and $12\arcsec$, 57~s
and 35~mK per 1~MHz channel for the (2--1) line.

The distribution of velocity-integrated CO-line intensities,
$I_{1-0}=\int T_\mathrm{mb}(CO_{1-0}) {\rm dv}$, is shown in
Fig.~\ref{fig:co-map}a. The r.m.s. noise in the velocity-integrated
emission varies between the fields, but is typically about $0.38\Kkms$
in the southern part and about $0.33\Kkms$ in the northern half of the
survey or $0.35\pm 0.10\Kkms$ for the total map (see
Table~\ref{tab:data}). We note that the sensitivity of our survey to
point-like and extended sources exceeds that of Loinard et al.
(\cite{loinard+99}) by factors $>8$ and $>1.5$, respectively.

%--------------------------
\section {The CO brightness distribution}

The survey of the Andromeda galaxy presented in this paper is the
largest and most detailed molecular-line survey ever made of an
extragalactic object. Most of the emission in Fig.~\ref{fig:co-map}
appears concentrated to radii between 3 and 12~kpc and occurs as long
and narrow filaments that strongly suggest a spiral arm structure. In
addition, we see a number of scattered CO clouds of weak intensity
between the spiral arms; sometimes these form bridges. The inner arms at
radii near 5~kpc are remarkably bright, especially in the NE half of the
galaxy.

Close to the centre of M\,31 CO emission is very weak. Melchior et al.
(\cite{melchior+00}) have found CO emission of about $T_{\rm A}^* =
20$~mK in a dark-cloud complex located at a distance of 350~pc from
the centre using position switching and long integrations. The
integration time per beam was about 6 hours and the r.m.s. noise 2~mK
per $3\kms$-wide channel, which is much better than the noise of the
present survey.

\begin{figure*}[htb]
\begin{center}
\includegraphics[bb = 26 183 551 798,width=12cm]{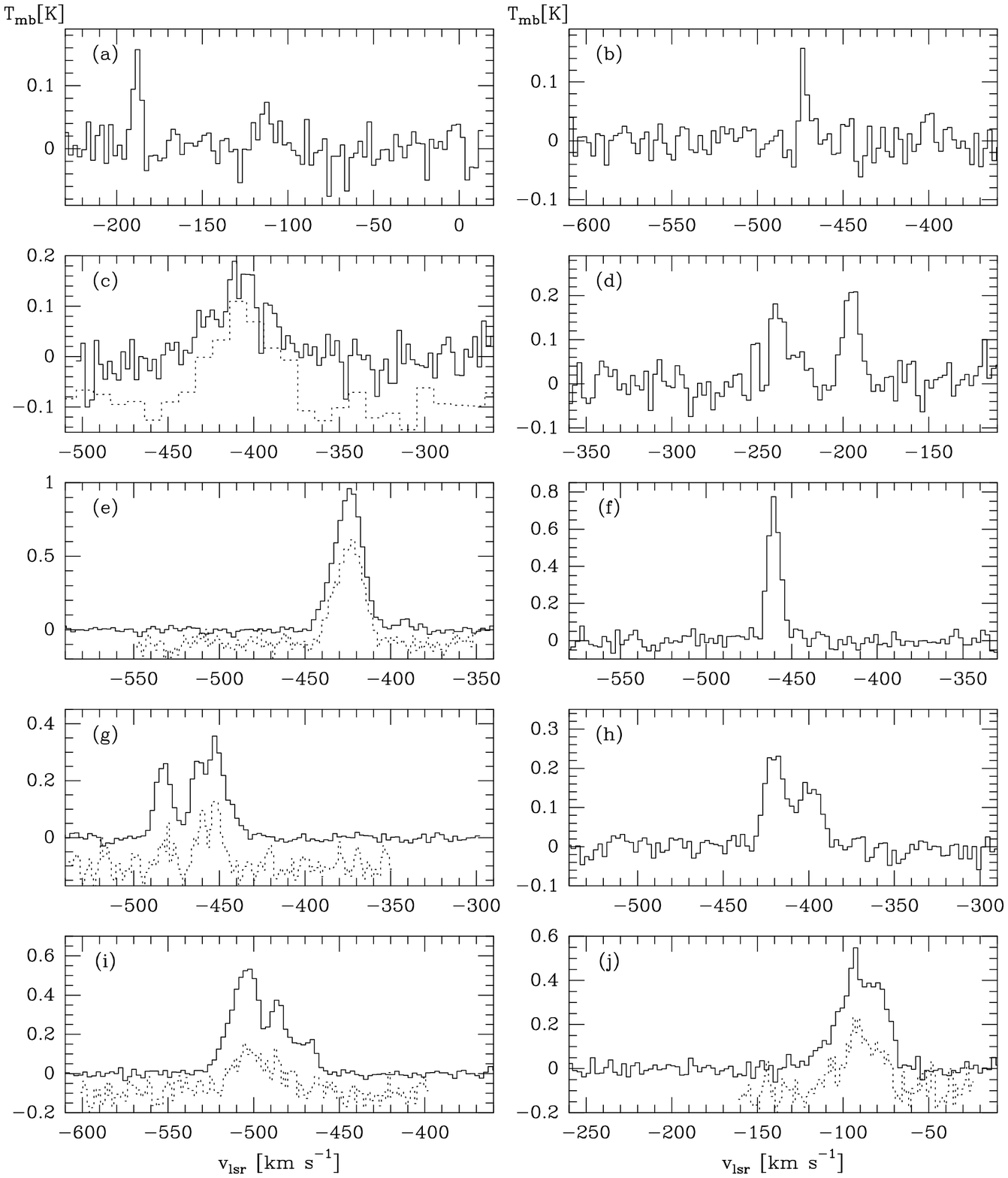}
\caption{ \label{fig:spectra}
Sample of spectra extracted from the data cube of the
\element[][12]{CO}(1--0) survey of M\,31 (full lines). At some positions
also the \element[][12]{CO}(2--1) spectrum is shown (dotted lines,
shifted down by 0.1~K for clarity). The velocity resolution is $2.6\kms$
for the (1--0) line and $1.3\kms$ for the (2--1) line. Velocities are
in the LSR frame.
{\bf (a)} Weak emission at $X,Y=46\farcm 1,\ -15\farcm 6$ which
corresponds to a deprojected distance from the centre of 19.4~kpc.
{\bf (b)} Spectrum with linewidth $<5\kms$ at $X,Y= -46\farcm 08,\
9\farcm 35$.
{\bf (c)} Broad line of about $40\kms$ close to the minor axis at
$X,Y = -0\farcm 33,\ -3\farcm 49$. The (2--1) spectrum is smoothed to
$9.9\kms$.
{\bf (d)} Two peaks separated by about $40\kms$ at $X,Y = 3\farcm
02,\  -4\farcm 97$.
{\bf (e)} Emission from the dust cloud D\,84 at $X,Y=-16\farcm
528,\ -8\farcm 728$; strongest emission on 10-kpc arm.
{\bf (f)} Emission at the position $X,Y=-17\farcm 8,\ -3\farcm 85$
near the dust cloud D\,153.
{\bf (g)} Emission with three velocity components near the dust
cloud D\,47 at $X,Y=-22\farcm 48,\ -7\farcm 53$.
{\bf (h)} Double peak emission from the inner arm, $X,Y= -7\farcm
57,\ 5\farcm 115$.
{\bf (i)} Emission with three components near the dust cloud D\,39,
$X,Y=-41\farcm 9,\ 8\farcm 54$.
{\bf (j)} Emission at the position $X,Y=24\farcm 85,\ 1\farcm 62$ near
the dust cloud D\,615. }
\end{center}
\end{figure*}

Only a few scattered clouds are visible at large distances from the
centre. The most distant cloud in this survey was found at a
deprojected radius of 19.4~kpc. The spectrum of this cloud is shown in
Fig.~\ref{fig:spectra}a. The central velocity of the emission fits well
to the velocity field of M\,31 (see Fig.~\ref{fig:co-map}b). The cloud
is located in the outermost part of the spiral arm A--N modelled by
Braun (\cite{braun91}) near several \ion{H}{ii} regions.

In the following subsections we model the spiral pattern and analyze
the arm-interarm contrast.

%Fig. 2

%%--------------------------
\subsection {The structure of molecular spiral arms}

Although M\,31 was classified from the beginning by Hubble
(\cite{hubble29}) as a {\it Sb} type spiral, attempts to draw its
spiral pattern have been mostly inconclusive. For example, Baade
(\cite{baade63}) used the young stars and Hodge (\cite{hodge79}) the
open star clusters as spiral arm tracers. The results have been
summarized by Hodge (\cite{hodge81b}). Due to absorption of optical
light by dust-rich lanes, the presence of a bright optical
bulge and the lack of \ion{H}{i} radio line emission in the inner disk,
there was no consensus even on whether the arms are trailing or
leading. More recently, Braun (\cite{braun91}) proposed a non-planar
trailing two-armed spiral pattern with varying pitch angle on the basis
of the \ion{H}{i} interferometric data. His model accounts fairly
well for the outer \ion{H}{i} arms, but it does not trace the innermost
structures. Based on their analysis of the $175\,\mu$m dust emission,
Haas et al. (\cite{haas+98}) suggested that M\,31 may be closer to a
ring galaxy than to a spiral.

The CO emission, which traces the dense molecular gas, is better
suited to determine M\,31's spiral pattern because the CO arms are
thinner than the \ion{H}{i} arms, they are less patchy than the
H$\alpha$ arms and not affected by absorption. As is discussed
below, the CO arm-interarm contrast is also much larger than in
\ion{H}{i}. Furthermore, the linear resolution of the CO survey is
sufficient to distinguish neighbouring arms even on the minor axis.

%Fig. 3
\begin{figure*}[htb]
\includegraphics[bb = 48 49 521 373,width=17cm,clip=]{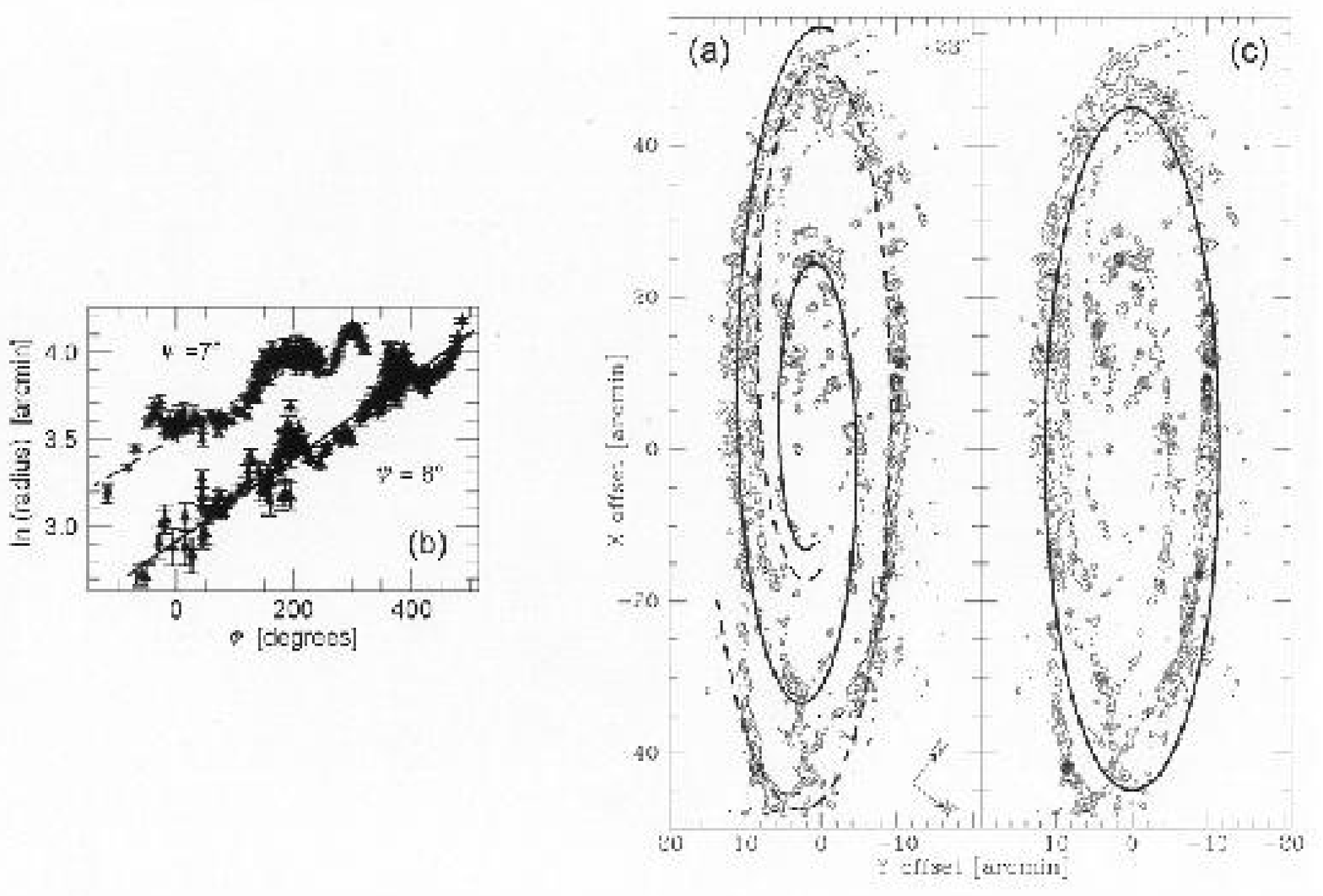}
\caption{ \label{fig:spiral}
Spiral structure in M\,31.
{\bf (a)} Two logarithmic spirals (solid and dashed lines) fitted
through the distribution of CO `clouds' in the radial range
$R=4$--12~kpc. They are overlayed onto the CO brightness distribution
(contours).
{\bf (b)}  Least-squares fits in the $\ln R-\phi$ plane
yielding pitch angles of $8\degr$ and $7\degr$ for the solid and dashed
spirals in (a), respectively. $\phi$ is the azimuthal angle in the plane
of M\,31 counted clockwise from the eastern minor axis.
{\bf (c)} Ring fitted to the distribution of `clouds' at $R=9$--12~kpc
(solid line) superposed onto the CO-brightness distribution (contours). }
\end{figure*}

In order to determine the spiral pattern in an objective way, we
decomposed the CO map into 170 individual `clouds',
using the CLUMPFIND analysis program (Williams et al.\
\cite{williams+94}). CLUMPFIND provided the position of `clouds',
deprojected them (assuming an inclination of the molecular disk of
$i=77\fdg 5$) and transformed them into polar coordinates. The
resulting data points, plotted in semi-logarithmic coordinates, were
then least-squares fitted with two straight lines representing two
simple logarithmic spirals (see Figs.~\ref{fig:spiral}a and b). The
fits led to very similar pitch angles for the two spirals, namely $\psi
= 8\degr\pm 1\degr$ and $7\degr\pm 1\degr$, and to a phase shift of
roughly $200\degr$. The derived pitch angles are in good agreement with
the value of $\psi = 6\fdg 7$ for the mean of the 2 spiral arms
observed in \ion{H}{i} (Braun\ \cite{braun91}) and with the
mean value of $7\fdg 4$ for the optical spiral arms (Baade \& Arp\
\cite{baade+arp64}). The oscillations visible in Fig.~\ref{fig:spiral}b
indicate variations in pitch angle which could be partly due to
variations in inclination angle observed by Braun (\cite{braun91}).

We note that these geometrical spiral fits required centre positions
shifted to the NE with respect to the nucleus of M\,31, i.e.
$X_\mathrm{c},Y_\mathrm{c} = 7\farcm 7,\  0\farcm 4$ for the full-line
spiral and $X_\mathrm{c},Y_\mathrm{c} = 5\farcm 8,\  2\farcm 3$ for the
dashed spiral (Figs.~\ref{fig:spiral}a and b). Such a displacement is
not surprising as the bulge of M\,31 is likely to host a bar (Stark \&
Binney\ \cite{stark+binney94}), which makes it difficult to trace the
spiral arms down to the nucleus.
This displacement and the fact that on the minor axis the outer CO
arms lie slightly outside the fitted spirals
may also indicate that the molecular spiral arms have
inclinations that differ from the adopted inclination of $77\fdg 5$ of
the main plane of M\,31, like the \ion{H}{i} arms (Braun\
\cite{braun91}). For a detailed comparison with the \ion{H}{i} arms a
more complete analysis of the molecular arm structure should be made,
which also includes the velocity structure of the arms. This could be
the subject of a future study.

For comparison, we have tried to represent the CO emission at
$R=8$--12~kpc by a circular ring, as suggested by Haas et al.
(\cite{haas+98}). The result of the fit is shown in
Fig.~\ref{fig:spiral}c. Even at these radii the spiral pattern appears
to be a better description of the CO distribution than a ring-like
structure.

The derived spiral pattern is a good fit to the CO emission up to about
12~kpc from the centre, except for the region of the giant stellar
association NGC~206 ($X,Y= -40\arcmin,\ +2\arcmin$), the velocity of
which strongly deviates from the normal rotation velocity.
In particular, it accounts for the splitting of the CO emission along
the eastern minor axis into two distinct arms. These arms are clearly
seen in the $8\,\mu$m emission from hot dust (MSX satellite map, Moshir
et al.\ \cite{moshir+99}; ISOCAM map of the central bulge of Willaime
et al.\ \cite{willaime+01}, see Fig.~\ref{fig:7} below) and are also
predicted by Braun's model although they are only marginally visible
in \ion{H}{i} and H$\alpha$ emissions.

At radii $\ga 12$~kpc the fitted spiral pattern starts to deviate from
the observed spiral. There the CO emission becomes fainter and the
filaments more difficult to trace. We note, however, that the ISO
$175\,\mu$m map of Haas et al. (\cite{haas+98}) and the H$\alpha$ image
of Walterbos (\cite{walterbos00}) show a clear arm-like structure that
seems to lie on the extension of the fitted spiral to the NE, far
beyond our CO image.

\subsection {Arm--interarm contrast}
\label{sec:contrast}

\begin{figure*}[htb]
\begin{center}
\includegraphics[bb = 33 64 492 681,angle=270,width=15cm,clip=]{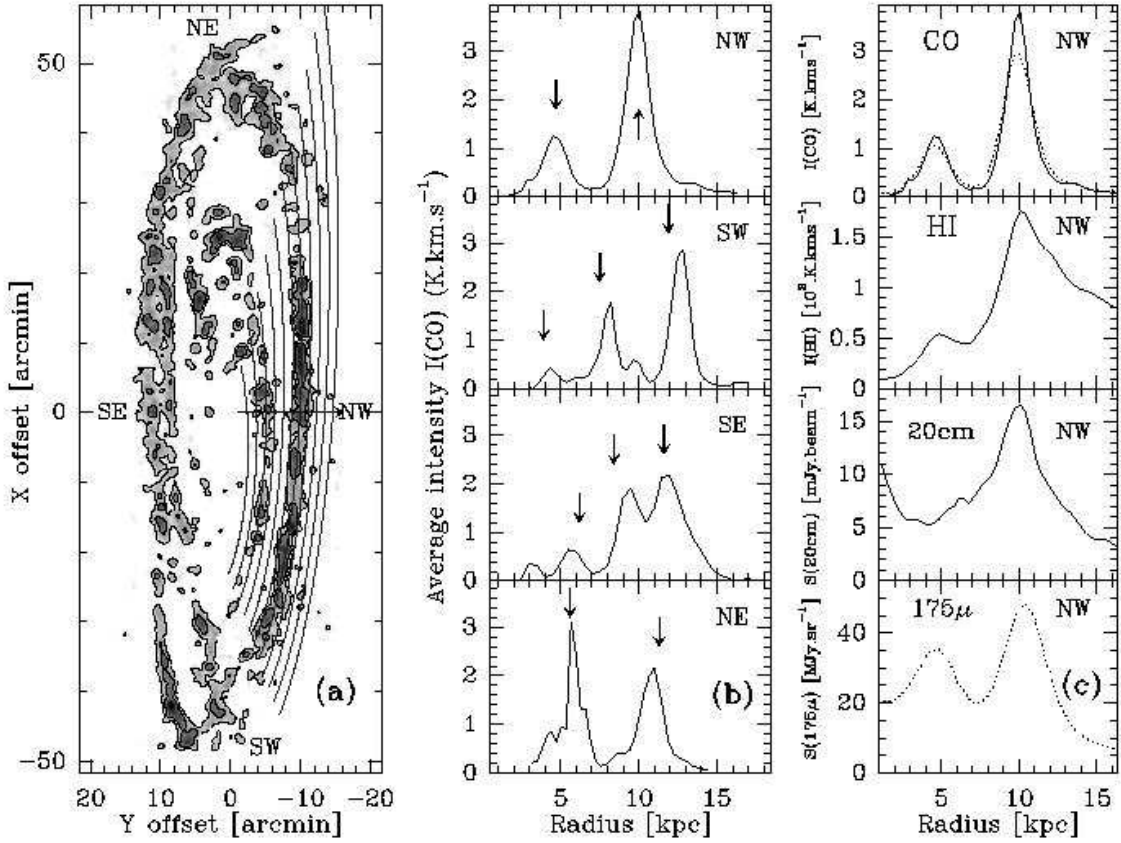}
\caption{\label{fig:contrast}
{\bf (a)} A family of logarithmic spiral segments (defined in the plane
of M\,31) fitting the CO arms in the NW quadrant, superimposed onto the
velocity-integrated CO emission, $I_{1-0}$, smoothed to a resolution of
$45\arcsec$. The spiral segments densely cover the NW quadrant between
$Y=1\arcmin$ and $Y= 14\arcmin$ on the minor axis (indicated by the
arrow).
{\bf (b)} {\it Top}: Mean profile of the CO intensity across the 5~kpc
and 10~kpc spiral arms, averaged along the spiral segments shown in
Fig.~\ref{fig:contrast}a.
{\it Middle and Bottom}: same as the upper figure for the SW, SE and NE
quadrants. The scale on the abscissa shows the galactocentric distance
along the minor axis (where $1\arcmin$ corresponds to 1050~pc) for the
NW and SE quadrants, and along the major axis for the SW and NE
quadrants (where $1\arcmin$ corresponds to 227~pc). The arrows
indicate the points of intersection of the two-armed logarithmic spiral
pattern with the minor or major axis; note that they do not always
coincide with the CO arms.
{\bf (c)} Comparison of the mean CO, HI, 20~cm continuum and
$175\,\mu$m continuum emission profiles in the NW quadrant. The
\ion{H}{i} data of Brinks \& Shane (\cite{brinks+shane84}) were
smoothed to an angular resolution of $45\arcsec$, the original
resolution of the 20~cm continuum data of Beck et al. (\cite{beck+98}).
The $175\,\mu$m continuum data of Haas et al. (\cite{haas+98}) have a
resolution of $90\arcsec$. The CO profiles in the upper figure were
obtained from the CO map smoothed to resolutions of $45\arcsec$ (full
line) and $90\arcsec$ (dotted line), respectively.
}
\end{center}
\end{figure*}

To derive the apparent arm width and the arm-interarm brightness
intensity ratio, we divided M\,31's disk into 4 quadrants of
approximately equal size. For each quadrant, we generated a family of
spiral segments from a linear combination of the spirals that best
fit the inner and outer CO arms. Figure~\ref{fig:contrast}a shows a sample
of the spiral segments generated for the NW quadrant, superposed onto
the CO map smoothed to a resolution of $45\arcsec$. We then calculated
the average CO intensity along each spiral segment.
Figure~\ref{fig:contrast}b shows these averages for the 4 quadrants as
functions of the spiral segment radii at their intersection with the
major or minor axis. In this analysis, we have discarded the highly
perturbed region surrounding NGC~206 in the SW quadrant.

The CO peak--to--dip ratio reaches a maximum $>20$ in all quadrants,
showing that the arm--interarm intensity contrast is high almost
everywhere. The shallow dip at $R= 11\kpc$ in the SE quadrant reflects
that the two brightest spiral arms ($R= 9\kpc$ and 12~kpc) are only
partly resolved in this quadrant (see Fig.~\ref{fig:co-map}). The
shoulders visible on Fig.~\ref{fig:contrast} at $R= 9\kpc$ (NE quadrant)
and at 10~kpc (SW quadrant) come from two weak, nearly symmetrical
arm-like substructures that are clearly visible on Fig.~\ref{fig:co-map}
(e.g from $X,Y= 27\arcmin,\, 8\arcmin$ to $X,Y= 40\arcmin,\, 2\arcmin$
in the NE quadrant).

That CO emission is very low between the arms, outside these
substructures, is supported by the more than one hundred pointed
observations we did in the interarm regions, in particular in the
direction of several weak and extended interarm features that appear
on the map of Loinard et al. (\cite{loinard+99}). Our observations
failed to reveal any emission down to levels 3--5 times lower than the
sensitivity of our OTF survey. Similarly, a strip of $8\arcmin$
length centred on $X,Y= 15\arcmin,\, 6\arcmin$, about half-way between
the 5~kpc and the 10~kpc arms did not show emission above $0.13\Kkms$ (1
s.d. at $23\arcsec$ resolution), which is 30 times below the average
intensity along the corresponding segment at the crest of the 10~kpc
arm.

The apparent half-power width of the arms on Fig.~\ref{fig:contrast}b
is 1--2 kpc. This includes the true arm width within the plane of M31,
the arm thickness perpendicular to the plane and the effect of
beam-smoothing. Nieten (\cite{nieten01}) attempted to disentangle these
3 effects for the NW part of the 10~kpc arm and derived a `true' width
of $500\pm100$~pc and an arm thickness of $150\pm 50$~pc.

%Fig. 4

We have repeated the above analysis for the \ion{H}{i} map of Brinks \&
Shane (\cite{brinks+shane84}), smoothed to the same resolution of
$45\arcsec$, for the 20~cm radio continuum map of Beck et al.
(\cite{beck+98}) at the original $45\arcsec$ resolution
and for the $175\,\mu$m map at $90\arcsec$ resolution (see
Fig.~\ref{fig:contrast}c). The arm--interarm contrast of the
integrated \ion{H}{i}-line brightness is $\simeq\,4$, which is 5 times
smaller than that of CO in the same disk section, and the apparent
half-power arm width of $6\arcmin$ is 3 times larger. Contrary to CO
emission, \ion{H}{i} emission is detected everywhere between the arms
at 5~kpc and 10~kpc. At $\lambda$20~cm the contrast is about 2.5 and the
arm width is in between the widths in CO and \ion{H}{i}. This was also
noted by Berkhuijsen et al. (\cite{elly+93}) at a resolution of
$75\arcsec$, who found that the width at $\lambda$20~cm corresponds to
the width of the total gas arm.

The molecular arms traced by CO are much narrower and thinner than the
\ion{H}{i} arms and the arm--interarm contrast is much higher. This
indicates that the molecular phase is short-lived compared to the life
time of the \ion{H}{i} gas.

From a comparison of CO emission and visual extinction in a wide strip
centred on the SW bright arm Neininger et al. (\cite{neininger+98})
concluded that CO(1--0) line emission is a good tracer of the molecular
gas in M\,31, including the interarm region. The $>5$ times larger
arm-interarm ratio in CO than in \ion{H}{i} thus implies that the
molecular gas has almost vanished in the interarm regions. Yet,
molecular and atomic clouds have about the same velocity dispersion (see
Sect.~5), have roughly the same response to the stellar potential and
follow the same orbits with the same orbital velocity. This indicates
that molecular clouds become mostly atomic when leaving the arms.

An accurate determination of the molecular cloud lifetimes would
require a density-wave model of M\,31 that explains the observed CO
arm pattern. Such a model is outside the scope of the present paper. A
crude upper limit to this lifetime can nevertheless be estimated from
the CO and \ion{H}{i} arm--interarm ratios. Consider a gas cloud
orbiting around M\,31's center and crossing a spiral arm. From
Fig.~\ref{fig:contrast}c we estimate that the total gas arm--interarm
ratio is 4--5. This means that, due to streaming motions, the clouds
stay 4--5 times longer inside the arms than they would if they followed
purely circular orbits at a constant velocity. For an arm width of
0.5~kpc, a pitch angle of $8\degr$ and an orbital velocity of
$300\kms$, the time spent in an arm is $5\, 10^7$~yr, which is an upper
limit to the age of the molecular clouds.

%--------------------------
\subsection {Line ratios in selected regions}

\begin{figure}[htb]
\includegraphics[bb = 138 37 562 793,width=8.8cm,clip=]{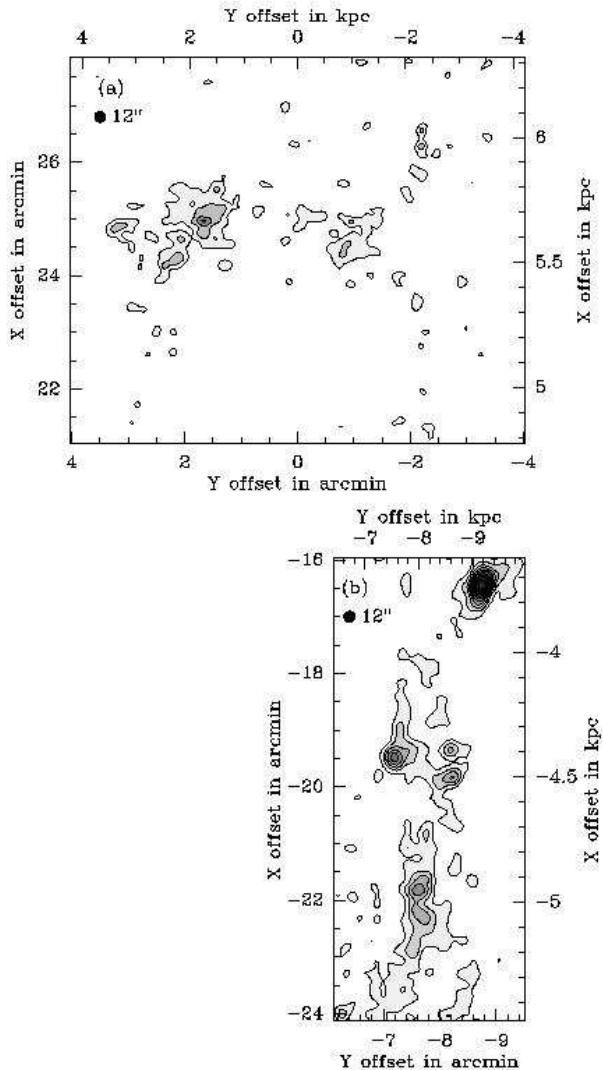}
\caption{\label{fig:mmaps}
Two maps showing the integrated emission of the \element[][12]{CO}(2--1)
transition at a resolution of $12\arcsec$. The contour lines correspond
to 3, 5, 7, \ldots times the noise ($T_\mathrm{mb}$ scale).
{\bf (a)} Spiral arm on the major axis at a distance of 5.5~kpc north
of the centre; the noise is $0.7\Kkms$.
{\bf (b)} Spiral arm region about 9~kpc west of the centre; the noise is
$1.2\Kkms$.}
\end{figure}

Our observations of several selected regions on a finer grid yielded
maps with high signal-to-noise ratio in the (2--1) line as well as in
(1--0). Two example maps of the velocity-integrated CO (2--1) line
intensity (Fig.~\ref{fig:mmaps}) cover bright segments, about 2--4 kpc
long, of spiral arms: one located near the major axis about 6~kpc from
the centre (Fig.~\ref{fig:mmaps}a), the other west of the main axis and
about 9~kpc from the centre (Fig.~\ref{fig:mmaps}b). The ratio $R_{21}$
of the (2--1)-to-(1--0) line intensities is often regarded as a first
indicator of excitation conditions in the clouds emitting the CO lines.
To apply this test to our data, we smoothed the (2--1) map to the same
resolution (23\arcsec ) as the (1--0) map, selected those points were
both line intensities were above $3\sigma$, and computed the line
ratio. In the region of Fig.~\ref{fig:mmaps}a we find an average ratio
of $0.65\pm 0.1$ with values reaching up to 1.0 near D615 ($X,Y =
25\farcm 5,\ 3\farcm 0$);\footnote{The dark cloud complexes (e.g.
D615) are indicated with their appellation in Hodge's Atlas of the
Andromeda Galaxy (Hodge\ \cite{hodge81a}).}
in the clouds of Fig.~\ref{fig:mmaps}b we find an average of $0.5\pm
0.1$ with ratios up to 0.7 around D84 ($X,Y
= -16\farcm 5,\ -8\farcm 7$).

%Fig. 5

These values are compatible with simple standard assumptions about the
CO excitation, i.e., that (i) both lines are optically thick, (ii) they
have the same excitation temperature $T_\mathrm{ex}$ which equals the
gas kinetic temperature $T_\mathrm{kin}$, and (iii) that they sample the
same gas of uniform conditions. The expected line ratio under these
conditions is $R_{21} = 0.49$ for $T = T_\mathrm{ex} = T_\mathrm{kin} =
3.5$~K, $R_{21} = 0.67$ for $T=6$~K, and $R_{21} = 0.79$ for $T=10$~K.
The line ratios we find in the spiral arms of M\,31 are also similar to
values found widely over the Milky Way (e.g., Sakamoto et al.\
\cite{sakamoto+97}) as well as those of 0.7 to 0.8 seen in M\,51
(Garcia-Burillo et al.\ \cite{garcia-burillo+93}) and 0.60 to 0.85 in
NGC\,891 (Garcia-Burillo et al.\ \cite{garcia-burillo+92}). For these
reasons the CO excitation in the bright regions of our survey is clearly
different from that of gas which shows very low ``subthermal'' values of
$R_{21} \sim 0.2$ to $\sim\,0.4$ found, e.g., by Allen \& Lequeux
(\cite{allen+lequeux93}) and Loinard et al. (\cite{loinard+95}) towards
some positions in M\,31. Of course these papers concern mostly positions
of small clouds or weak emission, whereas our values are selected for
positions with relatively strong lines and therefore probably high
ratios $R_{21}$. We note, however, that Melchior et al.
(\cite{melchior+00}) reported an $R_{21}$ ratio of 0.65 towards the
weak CO complex associated with D395A, located only 350~pc from the
centre of M\,31. This value is quite similar to the ratios that we find.
A more complete study of CO excitation in M\,31 would have to include
positions of weaker emission and also observations of optically thin CO
isotopes; this is beyond the scope of our present paper.

%Fig. 6
\begin{figure*}[p]
\includegraphics[bb = 71 109 507 709,angle=270,width=13.5cm,clip=]{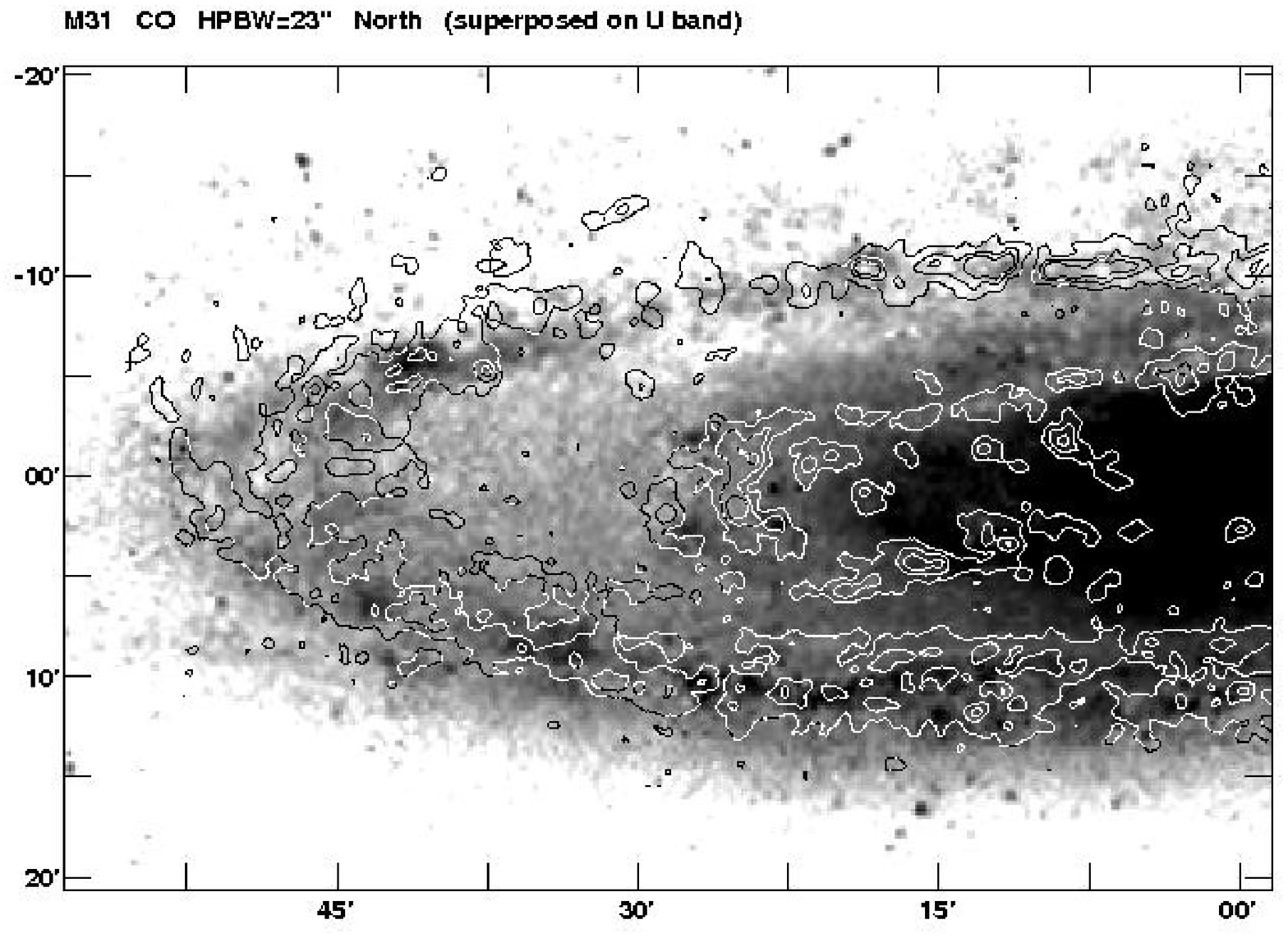}
\includegraphics[bb = 71 109 507 709,angle=270,width=13.5cm,clip=]{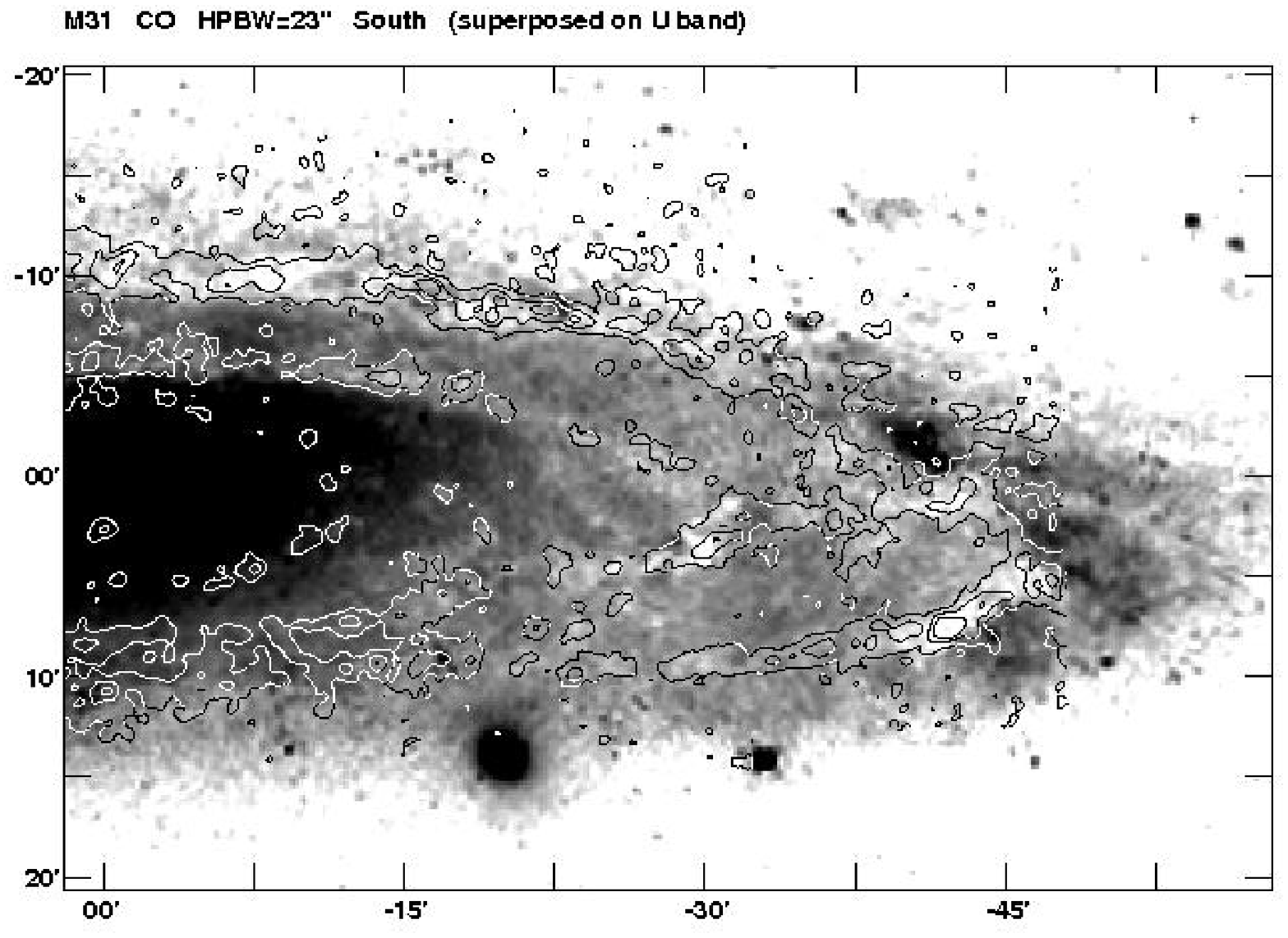}
\caption{\label{fig:6}
Contours of \element[][12]{CO}(1--0) emission from M\,31 at $23\arcsec$
resolution on a U-plate of Walterbos \& Kennicutt
(\cite{walterbos+kennicutt88}). The contour levels are 1
($=3\times$ r.m.s. noise), 4 and $8\Kkms$.
{\bf (a)} Northern half of M\,31.
{\bf (b)} Southern half of M\,31
}
\end{figure*}

\clearpage

\section {Comparison of CO with other gas and dust tracers}

\begin{figure}[htb]
\includegraphics[bb = 64 154 552 634,width=8.8cm,clip=]{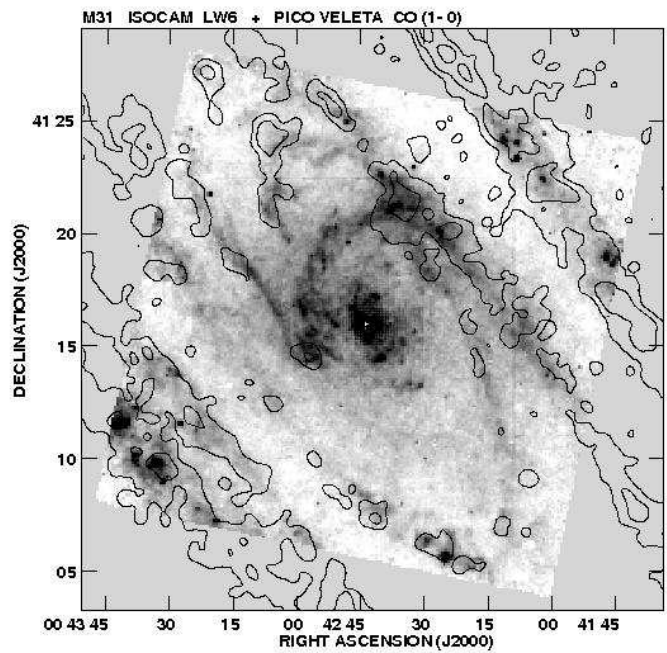}
\caption{\label{fig:7}
Contours of \element[][12]{CO}(1--0) emission from M\,31 at $23\arcsec$
resolution on the dust emission at $\lambda7.7\,\mu$m observed in the
central bulge (Willaime et al.\ \cite{willaime+01}, Fig. 2). The contour
levels are 1 ($=3\times$ r.m.s. noise), 4 and $10\Kkms$. The white dot
near the centre of the image indicates the centre of M\,31. Note the
alignment of CO clouds along the faint elliptical ring south
of the nucleus.
}
\end{figure}

The high resolution data presented in this paper allow detailed
comparisons with other tracers of the gas and the dust.
Figure~\ref{fig:6} shows
contours of CO brightness superimposed onto a U-band image
(Walterbos \& Kennicutt\ \cite{walterbos+kennicutt88}). The CO closely
traces the dust lanes, especially in the western part of M\,31 ($Y <
0\arcmin$) where the dust lanes stand out against the light of the
stellar bulge. The extended dark region around $X = 30\arcmin,\ Y =
-7\arcmin$ shows little CO, but this area has low brightness in many
constituents of M\,31 (see Fig.~\ref{fig:8}). The weak inner CO arm
crossing the major axis near $X= -12\arcmin$ appears to coincide with a
narrow dust lane seen at $7.7\,\mu$m by Willaime et al.
(\cite{willaime+01}) who observed the central part of M\,31 with ISOCAM
LW6 (see Fig.~\ref{fig:7}). The CO clouds near the major axis between
$X = 0\arcmin$ and $10\arcmin$ also coincide with dust features at
$7.7\,\mu$m, but not all dust features are seen in CO.

%Fig. 7

Figure~\ref{fig:8} collects the observed brightness distributions
of CO, \ion{H}{i}, total neutral gas, FIR at $175\,\mu$m and 20~cm radio
continuum. The $175\,\mu$m emission mainly traces cold dust at
temperatures near 16~K (Haas et al.\ \cite{haas+98}).
The difference between the distributions of CO and
\ion{H}{i} is striking. The \ion{H}{i} arms are much smoother and wider
than the CO arms and weak \ion{H}{i} emission is seen nearly everywhere
outside the spiral arms. We obtained the distribution of the total gas
column density, $N_\mathrm{gas} = N(\ion{H}{i}) + 2N(\mbox{H}_2)$, after
smoothing the CO map to the angular resolution of $24\arcsec\times
36\arcsec$ of the \ion{H}{i} map (Brinks \& Shane\
\cite{brinks+shane84}). We used the Galactic conversion factor
$X_\mathrm{CO} = N(\mbox{H}_2) / \int T_\mathrm{mb}
(^{12}\mbox{CO}_{1-0}) {\rm dv} = 1.9\times 10^{20}\, {\rm mol}\, {\rm
cm}^{-2}\, ({\rm K}\kms )^{-1}$ given by Strong \& Mattox
(\cite{strong+mattox96}), which was assumed to be constant across the
CO map. This conversion factor is supported by 1.2~mm observations of
the thermal emission from dust in two extended regions in the disk
(Zylka \& Gu\'elin, in prep.) and by virial mass estimates of several
molecular cloud complexes in M\,31 (Muller\ \cite{muller03}; Gu\'elin et
al.\ \cite{guelin+04}). The \ion{H}{i} column density was calculated
from the relation $N(\ion{H}{i})/\int T_\mathrm{b} {\rm dv} = 1.82\,
10^{18}\ {\rm at}\, {\rm cm}^{-2}$, valid for optically thin lines.
This relation may lead to a significant underestimate of
$N(\ion{H}{i})$ if $T_\mathrm{b} > 70$~K (Braun \& Walterbos\
\cite{braun+walterbos92}), i.e. at the crest of the arms. The
distribution of the total gas calculated this way is shown in
Fig.~\ref{fig:8}c.

Comparing the various distributions in Fig.~\ref{fig:8}, we notice that
the nuclear area is only prominent at $175\,\mu$m and $\lambda$20~cm
radio continuum, but all distributions in Fig.~\ref{fig:8} show
the pronounced ring of emission at about 10~kpc from the centre, where
also most of the star-formation regions are located. The spiral arm
closer to the nucleus, clearly visible in CO and at $175\,\mu$m, is
hardly seen in \ion{H}{i} and radio continuum. This indicates that the
CO--to--\ion{H}{i} brightness ratio increases towards smaller radii.
Only the \ion{H}{i} and the $175\,\mu$m distributions show extended
weak emission in between the spiral arms. Thus the CO emission is
concentrated to regions of the denser clouds seen at $175\,\mu$m and in
\ion{H}{i}, located in the spiral arms, and does not trace weak
and extended interarm emission at $175\,\mu$m, which is especially
visible at $Y> 0\arcmin$. In Sect.~4.2 we present first results of a
correlation study between CO, \ion{H}{i} and $175\,\mu$m. Radial
distributions and the gas--to--dust ratios are shown in Sect.~4.1.

The radio continuum emission at $\lambda$20~cm (Beck et al.\
\cite{beck+98}) is also concentrated in the main spiral arms that form
the emission ring. As at this wavelength most of the emission is
nonthermal, magnetic fields and cosmic rays are concentrated in this
ring. Berkhuijsen (\cite{elly97}), using the CO data of Koper et al.
(\cite{koper+91}), compared nonthermal emission with CO and \ion{H}{i}
data at a resolution of $9\arcmin$. Although significant correlations
indeed exist at this resolution, detailed correlations with the new
data are required to enable an interpretation in terms of the coupling
between magnetic fields and gas. This will be the subject of a
forthcoming study.

\subsection{Radial distributions}
In Fig.~\ref{fig:9} we show the radial variations of the observed CO and
\ion{H}{i} brightness distributions averaged in 230~pc-wide circular
rings in the plane of M\,31 (i.e. $1\arcmin$ on the major axis).
Because the spiral arms in M\,31 are tightly wound, the narrow CO arms
clearly stand out in the radial distributions. In the northern part ($X
> 0\arcmin$), the average intensity of the inner arm at $R \simeq
25\arcmin$ is comparable to that of the arms forming the bright
emission ring at $R\simeq 50\arcmin$, whereas in the southern part ($X
< 0\arcmin$) this inner arm is much weaker. The radial profiles of
\ion{H}{i} are much wider than those of CO, but the inner arm and main
ring are still indicated. The \ion{H}{i} disk extends as far as
$R\simeq 150\arcmin$, 34~kpc from the nucleus (Emerson\
\cite{emerson74}), about 2 times further than the molecular gas in the
map of Koper et al. (\cite{koper+91}).

%Fig. 8
\begin{figure*}[htb]
\includegraphics[bb = 79 57 515 717,width=14cm,clip=]{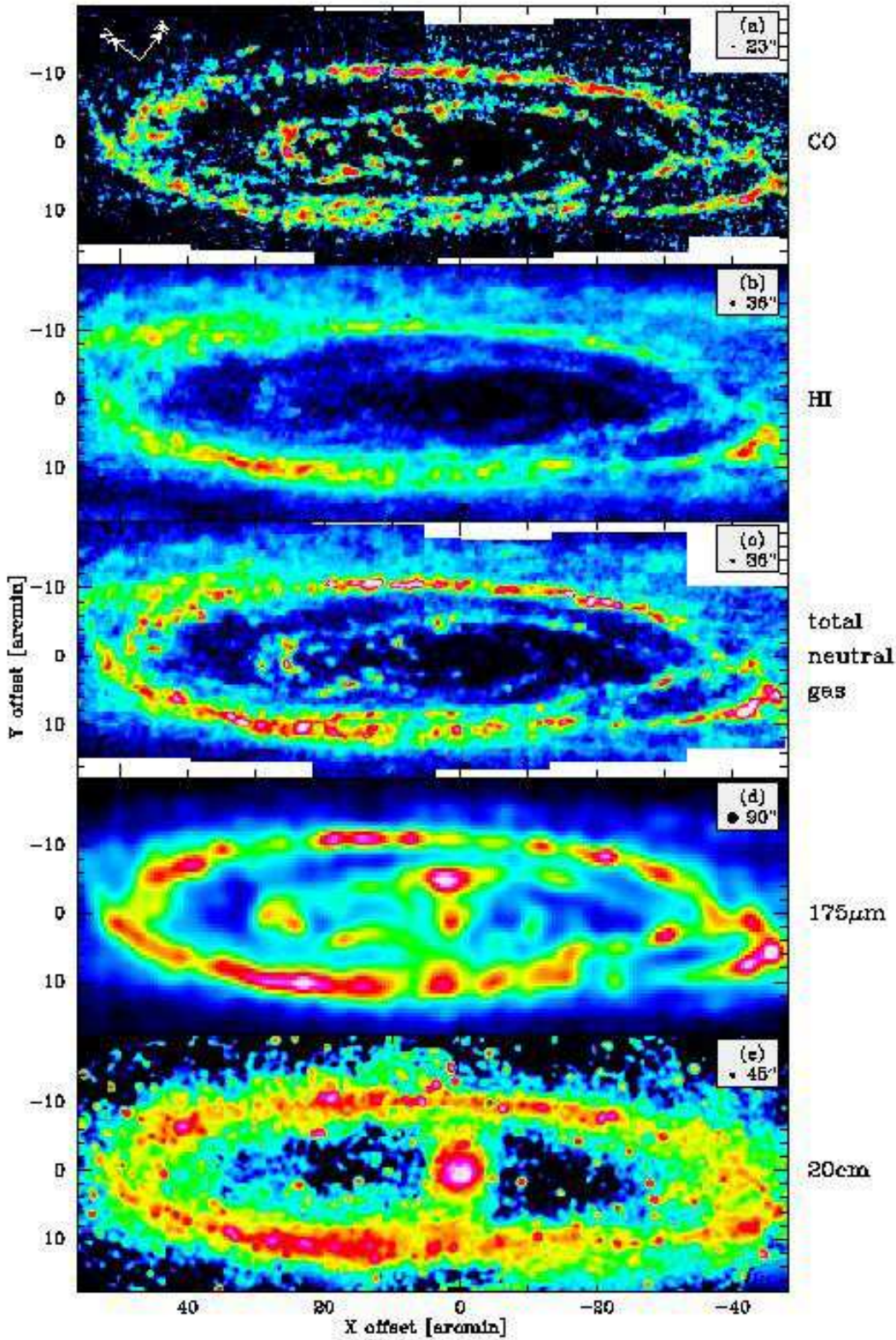}
\caption{\label{fig:8}
Distributions of neutral gas, cold dust and radio continuum in M\,31.
From top to bottom:
{\bf (a)} emission observed in the \element[][12]{CO}(1--0) line (this
paper),
{\bf (b)} emission observed in the \ion{H}{i} line (Brinks \& Shane\
\cite{brinks+shane84}),
{\bf (c)} emission from the total neutral gas, N(\ion{H}{i})+2N(H$_2$),
with $X_{\rm CO} = 1.9\times 10^{20}\, {\rm mol}\, {\rm cm}^{-2}\,
({\rm K}\kms )^{-1}$ (see text),
{\bf (d)} emission from cold dust at $\lambda175\,\mu$m (Haas et al.\
\cite{haas+98}),
{\bf (e)} radio continuum emission at $\lambda$20~cm (Beck et al.\
\cite{beck+98}). The half-power beamwidth is indicated in the upper
right-hand corner of each map.
The intensities at the maximum near $X,Y = -42\arcmin,\ 8\arcmin$
(white-red) and the minimum near $X,Y = -30\arcmin,\ 10\arcmin$ (blue-green)
in each map are: (a) 6.7 and 0.8 in $\rm 10^{21}\, \mbox{at}\,
\mbox{cm}^{-2}$, (b) 6.5 and 2.0 in $\rm 10^{21}\, \mbox{at}\,
\mbox{cm}^{-2}$, (c) 12.1 and 2.6 in $\rm 10^{21}\, \mbox{at}\,
\mbox{cm}^{-2}$, (d) 85 and 27 in MJy/sr, (e) 2.9 and 0.7 in
Jy/beam area. The M\,31 coordinates are based on the
centre position $\rm (\alpha ,\delta)_{50} = (0^h 40^m 00\fs 3,\
41\degr 00\arcmin 03\arcsec$) and a position angle of the major axis
$PA = 37\fdg 7$ (see footnote 1).}
\end{figure*}

\clearpage

Our CO profiles of the southern half are only half as strong
as those shown in Fig.~10d of Loinard et al. (\cite{loinard+99}),
whereas the \ion{H}{i} profiles are nearly identical. This may indicate a
difference in the adopted CO intensity scales, e.g. in the calibration
or in the corrections for beam efficiencies. We note that in Sect.~4.3
we find a close agreement of our CO intensity scale with that
of the 1.2-m telescope at the Center for Astrophysics (Dame et al.\
\cite{dame+93}), to which the standard Galactic $X_\mathrm{CO}$ is
calibrated. We also find that even after smoothing our map to the
effective resolution of $1\arcmin$ of the map of Loinard et al.,
the half-intensity width of the CO arms in their Figs. 5 and 6a is
about 1.5 times larger than in our map. We may attribute this
discrepancy to the much higher error beam of the FCRAO telescope
compared to that of the IRAM 30-m dish.

%Fig. 9
\begin{figure}[htb]
\includegraphics[bb = 96 55 557 691,width=8cm]{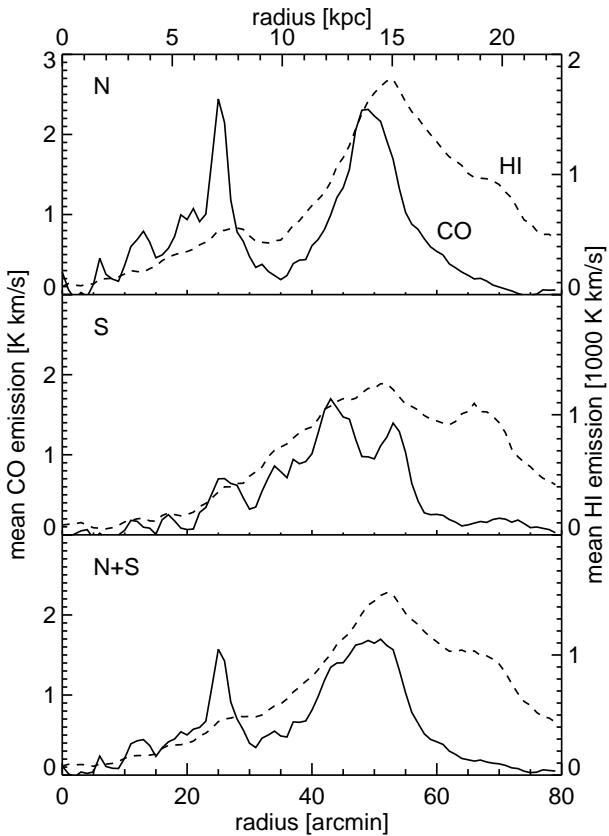}
\caption{\label{fig:9}
Radial profiles of $I_{1-0}$ (full lines, left-hand scale) and
$I_{\ion{H}{i}}$ (dashed lines, right-hand scale) for the northern half
(N), the southern half (S) and the full area of M\,31 (N+S). The
profiles show intensities along the line of sight averaged in circular
rings of $1\arcmin$ width in the plane of M\,31; typical standard
deviations are $0.01\Kkms$ and $5\Kkms$ for the CO and the \ion{H}{i}
profiles, respectively. The CO profiles were obtained from the map at
$23\arcsec$ resolution (Fig.~\ref{fig:co-map}) and the \ion{H}{i}
profiles from the map of B\&S at $24\arcsec \times 36\arcsec$
resolution. The limited extent of the CO map on the major axis, $R=
55\arcmin$ in the north and $R=50\arcmin$ in the south, reduces the
intensities of the profiles of $I_{1-0}$ beyond these radii by a few
percent. }
\end{figure}

%Fig. 10
\begin{figure}[htb]
\includegraphics[bb = 109 109 522 731,width=8cm,clip=]{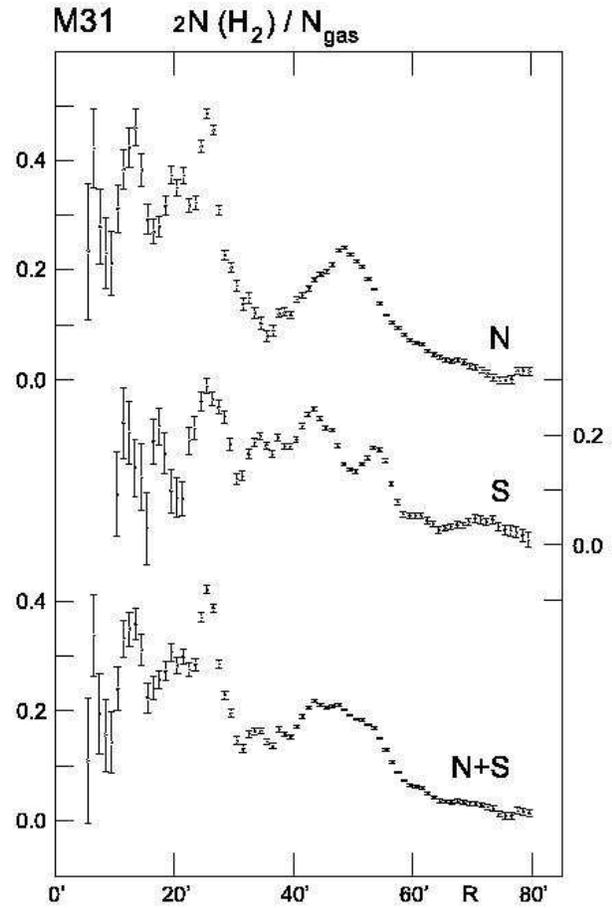}
\caption{\label{fig:10}
Radial variation of the molecular gas fraction
$2N(\mbox{H}_2)/N_\mathrm{gas}$ for the northern half (N, upper
left-hand scale), the southern half (S, right-hand scale) and the full
area of M\,31 (N+S, lower left-hand scale), where $N_\mathrm{gas} =
N(\ion{H}{i}) + 2N(\mbox{H}_2)$ and $X_\mathrm{CO} = 1.9\times
10^{20}\, \mbox{mol\, cm}^{-2}$ $\rm (K\, km\, s^{-1})^{-1}$ (Strong \&
Mattox\ \cite{strong+mattox96}). The CO map was smoothed to the
resolution of $24\arcsec \times 36\arcsec$ of the \ion{H}{i} map before
the molecular fractions were computed. The molecular
fractions were calculated from the column densities averaged in circular
rings of $1\arcmin$ width in the plane of M\,31.
}
\end{figure}

\begin{figure*}[htb]
\includegraphics[bb = 161 94 432 724,angle=270,width=18cm,clip=]{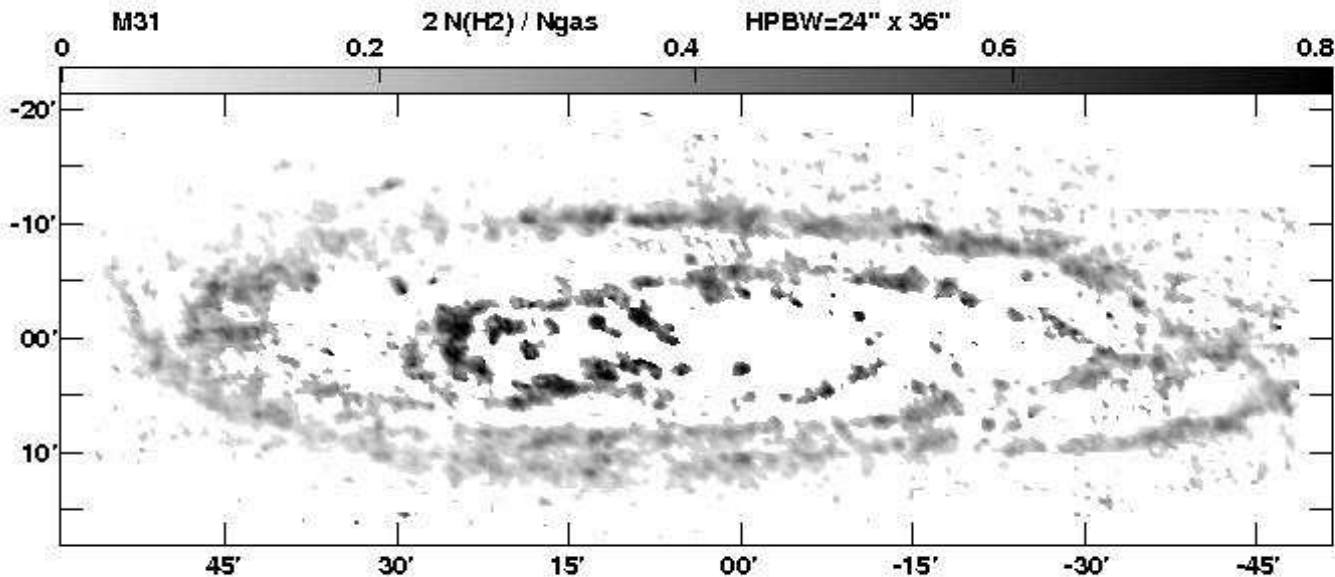}
\caption{\label{fig:11}
Distribution of the molecular gas fraction
$2N(\mbox{H}_2)/N_\mathrm{gas}$ in M\,31 at a resolution of $24\arcsec
\times 36\arcsec$. Only data points larger than $3\times$ r.m.s. noise
in the CO and \ion{H}{i} maps were used. $N_\mathrm{gas} =
N(\ion{H}{i}) + 2N(\mbox{H}_2)$ and a constant value of $X_\mathrm{CO}
= 1.9\times 10^{20}\, \mbox{mol\, cm}^{-2}$ $\rm (K\, km\, s^{-1})^{-1}$
(Strong \& Mattox\ \cite{strong+mattox96}) were assumed.
}
\end{figure*}

The fraction of molecular--to--total gas,
$2N(\mbox{H}_2)/N_\mathrm{gas}$, generally decreases with increasing
distance from the centre (see Figs.~\ref{fig:10} and \ref{fig:11}). The
highest values occur on the weak inner arm at $R\simeq 12\arcmin$
(2--3~kpc) and on the bright inner arm at $R\simeq 25\arcmin$ (5--6~kpc)
where the molecular fraction is nearly 0.5 in the north and about 0.25
in the south (see Fig.~\ref{fig:10}). On the bright ring at $R \simeq
50\arcmin$ the molecular fraction is about 0.2 both in the northern and
in the southern part, although the northern part of the ring contains
about 1.5 times more neutral gas than the southern one.
The averages in circular rings underestimate the ratios on the arms.
Figs.~\ref{fig:contrast}c and \ref{fig:11} yield a decrease from $\sim
0.6$ on the arms at 5~kpc to 0.3--0.4 on the arms at 10~kpc.
The observed values of
$2N(\mbox{H}_2)/N_\mathrm{gas}$ and its radial decrease are typical for
nearby galaxies (Honma et al.\ \cite{honma+95}). The decrease of the
molecular fraction in Fig.~\ref{fig:10} confirms the decrease in the
fraction of molecular gas mass along the major axis of M\,31 reported by
Dame et al. (\cite{dame+93}). Along the arms the molecular fraction
varies considerably (see Fig.~\ref{fig:11}). The highest fraction
detected is 0.96 for a cloud near the northern major axis at $X
= 13\arcmin,\  Y = -1\arcmin$, which seems purely molecular.

%Fig. 11
%Fig. 12
\begin{figure}[h]
\includegraphics[bb = 16 84 581 770,width=8.5cm]{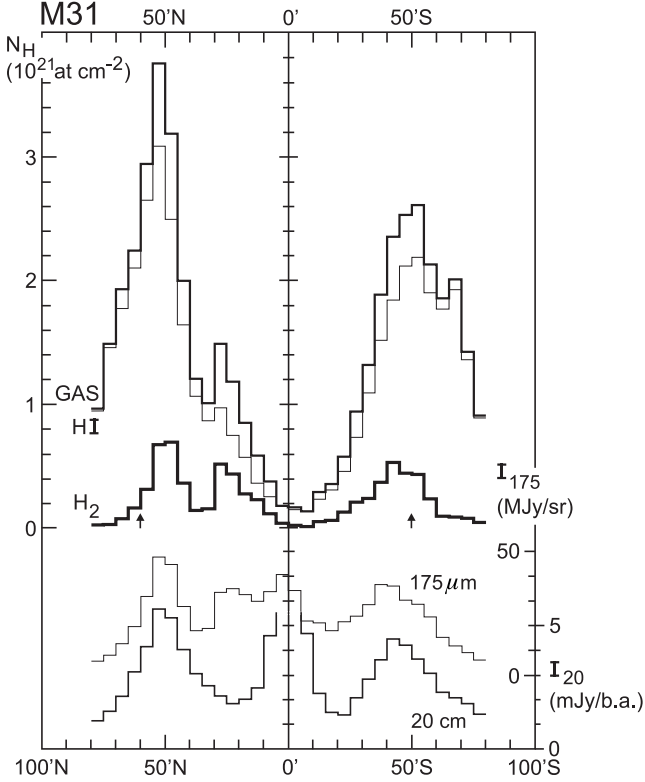}
\caption{\label{fig:12}
Radial variations for the northern (left panel) and the southern (right
panel) half of M\,31 of $N_\mathrm{gas}$, $N(\ion{H}{i})$ and
$2N(\mbox{H}_2)$ (left-hand scale) as well as of $I_\mathrm{175\mu m}$
(inner right-hand scale) and $I$(20\,cm) (outer right-hand scale)
obtained from the distributions in Fig.~\ref{fig:8}. The gas maps and
the radio continuum map were smoothed to the resolution of $90\arcsec$
of the $175\,\mu$m map. The line-of-sight surface brighnesses were
averaged in circular rings around the centre of $5\arcmin$ width in the
plane of M\,31. The two arrows indicate the extent of the CO map on the
major axis. }
\end{figure}

We smoothed the distributions of CO, \ion{H}{i}, total gas and 20~cm
radio continuum in Fig.~\ref{fig:8} to the angular resolution of
$90\arcsec$ of the $175\,\mu$m map and compare their radial
distributions in Fig.~\ref{fig:12}. In all constituents the main ring at
$R\simeq 50\arcmin$ is brighter in the north than in the south. The
pronounced molecular arm in the north at $R\simeq 25\arcmin$ is
invisible in radio continuum, possibly because of a lack of relativistic
electrons (Moss et al.\ \cite{moss+98}). Disregarding the central region
$R< 10\arcmin$, the profiles of molecular gas and $175\,\mu$m emission
from cold dust are most alike.

\begin{figure}[htb]
\includegraphics[bb = 157 93 688 496,width=8.8cm]{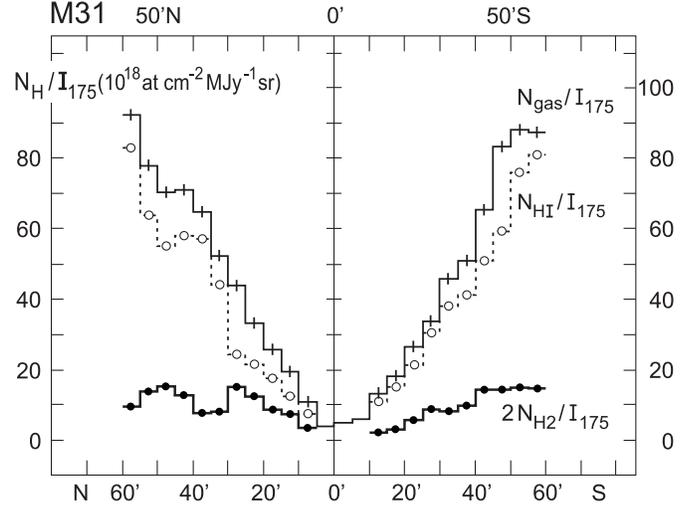}
\caption{\label{fig:13}
Radial variation of apparent
gas--to--dust ratios for the northern (left panel)
and the southern half (right panel) of M\,31 derived from the profiles
in Fig.~\ref{fig:12}. Thin full line -- $N_\mathrm{gas}/I_{175}$, dashed
line -- $N(\ion{H}{i})/I_{175}$ and thick full line --
$2N({\rm H}_2)/I_{175}$. $N_\mathrm{gas} = N(\ion{H}{i}) +
2N({\rm H}_2)$ and a constant value of $X_\mathrm{CO} = 1.9\times
10^{20}\, \mbox{mol\, cm}^{-2}$ $\rm (K\, km\, s^{-1})^{-1}$ (Strong
\& Mattox\ \cite{strong+mattox96}) were used. Note the increase
of $2N({\rm H}_2)/I_{175}$ on the molecular arm at $R\simeq 25\arcmin$
and on the main emission ring at $R\simeq 50\arcmin$.
}
\end{figure}

Using the profiles in Fig.~\ref{fig:12} we calculated the apparent
gas--to--dust ratios $2N({\rm H}_2)/I_{175}$, $N(\ion{H}{i})/I_{175}$
and $N_\mathrm{gas}/I_{175}$ as a function of radius which are presented in
Fig.~\ref{fig:13}. The ratio $2N({\rm H}_2)/I_{175}$ is clearly
enhanced in the spiral arms at $R\simeq 25\arcmin$ and $R\simeq
50\arcmin$, especially in the north, whereas $N(\ion{H}{i})/I_{175}$
continuously increases steeply from the centre outwards by nearly a
factor 20. As \ion{H}{i} is the dominant gas component, the apparent
total gas--to--dust ratio $N_\mathrm{gas}/I_{175}$ increases by
about a factor 20 between $R=0\arcmin$ and $R=60\arcmin$.
However, the physical reality of this strong increase may be
questionable. Firstly, the \ion{H}{i} line opacity changes with radius;
secondly the $X_\mathrm{CO}$ conversion factor may vary with radius;
thirdly, $I_{175}$ may not reflect the dust column density if the dust
absorption cross section and, especially, the dust temperature vary
with radius.

%Fig. 13

The first two causes are unlikely. Braun \& Walterbos
(\cite{braun+walterbos92}) showed that the atomic gas temperature in
M\,31 increases towards larger radii, while the 21-cm line opacity
decreases. This variation, however, is less than $\sim20\%$ and cannot
account for the strong gradient in $N(\ion{H}{i})/I_{175}$.

The behaviour of $X_\mathrm{CO}$ with radius has been the subject of
several studies in the Milky Way and nearby galaxies (Wilson\
\cite{wilson95}; Sodroski et al.\ \cite{sodroski+95}; Strong et al.\
\cite{strong+04}). $X_\mathrm{CO}$ is found to increase with increasing
radius, perhaps in relation with decreasing metallicity. Such an
increase, if present at all in M\,31, would only enhance the radial
variations in $N({\rm H}_2)/I_{175}$ and $N_\mathrm{gas}/I_{175}$.

A study of 50 bright cloud complexes at a resolution of at
least $12\arcsec$, which is high enough to alleviate rotation velocity
gradients, at radial distances between 5 and 12~kpc (Muller\
\cite{muller03}; Gu\'elin et al.\ \cite{guelin+04}) yield
$X_\mathrm{CO}= (1-5)\, 10^{20}\, \mbox{mol\, cm}^{-2}$ $\rm (K\, km\,
s^{-1})^{-1}$ without a  radial dependence.
Melchior et al. (\cite{melchior+00}) and Israel et al.
(\cite{israel+98}) find similar values for two complexes at $R=
0.35\kpc$ and $R=2.5\kpc$. Hence, except perhaps for the observation by
Allen \& Lequeux (\cite{allen+lequeux93}) of a dark cloud (D268,
$R=2.5\kpc$) with low CO luminosity and large
linewidth, which may not be in equilibrium, there is no evidence
of a radial variation of $X_\mathrm{CO}$ in M\,31.

It is more difficult to rule out a variation of the dust temperature
$T_\mathrm{d}$ with radius. The thermal emission of cold dust at
$175\,\mu$m varies like $T_\mathrm{d}$ to the power 4--5, so even a mild
decrease of $T_\mathrm{d}$ outwards could cause a strong decrease of
$I_{175}$. In fact, the apparent increase of $N_\mathrm{gas}/I_{175}$
from 1 to 14~kpc could be explained by a decrease of the dust
temperature from $\sim25$~K to 15~K. Only extensive mapping of this
emission
at longer wavelengths will allow to discriminate between a temperature
effect and a dust--to--gas variation.

\begin{figure*}[htb]
\includegraphics[bb = 144 87 453 747,angle=270,width=18cm]{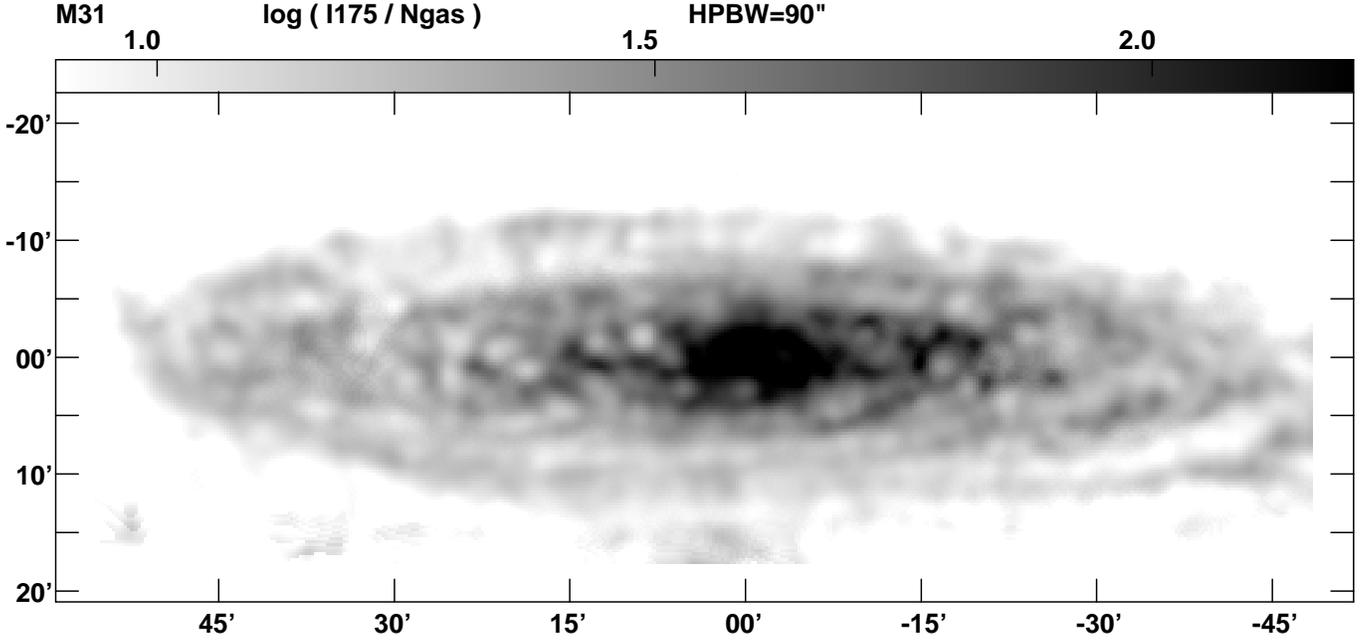}
\caption{\label{fig:14}
Distribution of the apparent
dust--to--total gas ratio, $I_{175}/N_\mathrm{gas}$
$\rm (MJy\, sr^{-1}) / (10^{21}\, at\, cm^{-2})$, in M\,31 at
$90\arcsec$ resolution. Local minima (holes) are caused by enhancements
in $N({\rm H}_2)$ (compare Fig.~\ref{fig:co-map}a or Fig.~\ref{fig:11}). }
\end{figure*}

An increase of the apparent gas--to--dust ratio with increasing radius in
M\,31 was found by several authors from comparisons of $N(\ion{H}{i})$
and optical or UV extinction (Bajaja \& Gergely\
\cite{bajaja+gergely77}; Walterbos \& Kennicutt\
\cite{walterbos+kennicutt88}; Xu \& Helou\ \cite{xu+helou96};
Nedialkov et al.\ \cite{nedialkov+00}; Savcheva \& Tassev\
\cite{savcheva+tassev02}). The latter authors used globular clusters
and found a similarly strong increase of the apparent
gas--to--dust ratio as we do. The distribution across M\,31 of the
reversed ratio, i.e. the apparent dust--to--total gas ratio
$I_{175}/N_\mathrm{gas}$, is shown in
Fig.~\ref{fig:14}. The general decrease from the centre outwards is
clearly visible. $I_{175}/N_\mathrm{gas}$ tends to be low on the gas
arms where the CO emission is strong. Even the many roundish minima in
the distribution of $I_{175}/N_\mathrm{gas}$ coincide with CO clouds.

It is interesting to note that on the brightest spot of emission at
$175\,\mu$m, about $4\arcmin$ west of the nucleus, the value of
$I_{175}/N_\mathrm{gas}$ is normal for that radius. This bright emission
comes from one of the coldest dust clouds in M\,31 analyzed by
Schmidtobreick et al. (\cite{schmidto+00}). Willaime et al.
(\cite{willaime+01}) showed that this cloud is located at a junction or
superposition of several thin dust lanes seen at $7.7\,\mu$m (see
Fig.~\ref{fig:7}), which may explain its normal dust--to--gas ratio.

%Fig. 14

%Table 2
\begin{table*}[htb]
%\begin{center}
\caption{Correlations between CO, \ion{H}{i} and $175\,\mu$m}
\label{tab:correl}
\begin{tabular}{llcr@{$\pm$}lr@{$\pm$}lr@{$\pm$}lc}
\hline\hline
\noalign{\smallskip}
$X$    &$Y$  &$\Delta R$  &\multicolumn{4}{c}{$\log Y=a\log X+b^{1)}$}
       &\multicolumn{2}{c}{Correl.} &N$^{1)}$\\
       & & (\arcmin ) &\multicolumn{2}{c}{a} &\multicolumn{2}{c}{b}
       &\multicolumn{2}{c}{coeff. $\rho$}\\
\noalign{\smallskip}
\hline\noalign{\smallskip}
$I_{\ion{H}{i}}$  &$I_{1-0}$ &\pheins 0--35 &2&5 &$-$5&20
     &0.3&0.1 &\pheins 90\\
$\rm (K\kms )$  &$\rm (K\kms )$ &35--60 &1.7&1.0 &$-$5.2&2.4
     &0.61&0.04 &234\\
     &   &\pheins 0--60 &1.3&0.5 &$-$4.1&1.4 &0.50&0.04 &324\\
\noalign{\smallskip}
$I_{175}$    &$I_{1-0}$ &\pheins 0--35  &1.6&1.1 &$-$2.53&0.63
     &0.56&0.07 &\pheins 94\\
$\rm (MJy\, sr^{-1})$  &$\rm (K\, km\, s^{-1})$ &35--60 &1.57&0.20
     &$-$2.39&0.16 &0.81&0.02 &234\\
     & &\pheins 0--60  &1.60&0.22 &$-$2.47&0.16 &0.75&0.02 &328\\
\noalign{\smallskip}
$I_{175}$    &$I_{\ion{H}{i}}$  &\pheins 0--35  &0.8&1.8 &1.4&2.2
     &0.51&0.07 &125\\
$\rm (MJy\, sr^{-1})$  &$\rm (K\, km\, s^{-1})$ &35--60 &0.84&0.27
     &1.85&0.44 &0.75&0.03 &264\\
     & &\pheins 0--60 &1.09&0.32 &1.36&0.42 &0.54&0.04 &388\\
\noalign{\smallskip}
$I_{175}$    &$N_{\rm gas}$ &\pheins 0--35   &1.42&0.53 &0.89&0.60
     &0.68&0.05 &134\\
$\rm (MJy\, sr^{-1})$  &$\rm (10^{18}\, at\, cm^{-2})$ &35--60
     &1.12&0.14 &1.64&0.22 &0.85&0.02 &263\\
     & &\pheins 0--60 &1.23&0.19 &1.27&0.27 &0.67&0.03 &395\\
\noalign{\smallskip}
\hline
\noalign{\smallskip}
\multicolumn{10}{l}{$^{1)}$Weighted fits of the bisector through N
pairs of (log\,$X$, log\,$Y$), where N is the number}\\
\multicolumn{10}{l}{\enskip of independent points (Isobe et al.\
\cite{isobe+90}; Nieten\ \cite{nieten01})}
\end{tabular}
%\end{center}
\end{table*}

\subsection{Correlations between the distributions of CO,
\ion{H}{i} and FIR}

\begin{figure*}[p]
\includegraphics[bb = 41 27 522 762,width=13.5cm,clip=]{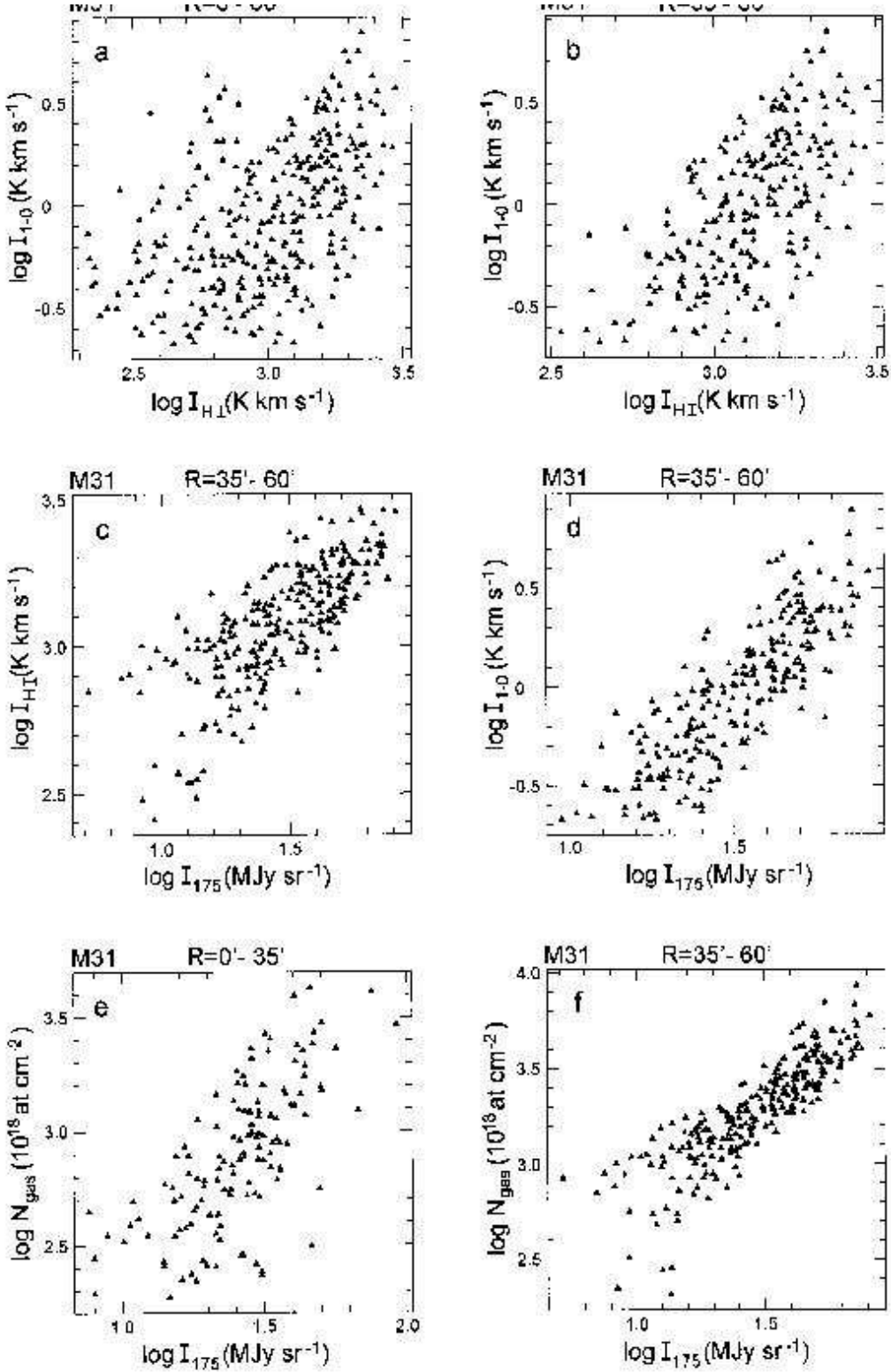}
\caption{\label{fig:15}
Examples of correlations between neutral gas and cold dust (16\,K) at
$90\arcsec$ resolution for two radial ranges in M\,31. Only independent
points ($1.67\times$ HPBW apart) above $3\times$ r.m.s. noise are
used, in panels (a), (b), (d) determined by the noise in the CO map and
in (c), (e), (f) by that in the \ion{H}{i} map. The resulting fits of
the bisector are given in Table~\ref{tab:correl}.
{\bf (a)} $I_{1-0}$ as a function of $I_{\ion{H}{i}}$ for $R =
0\arcmin -60\arcmin$, and
{\bf (b)} for $R = 35\arcmin -60\arcmin$. The secondary maximum in
(a) at $\log I_{\ion{H}{i}} \simeq 2.8$ is due to the inner arm
at $R\simeq 25\arcmin$.
{\bf (c)} $I_{\ion{H}{i}}$ as a function of $I_{175}$ for $R =
35\arcmin -60\arcmin$. The two branches visible at $\log I_{175} < 1.3$
result from the strong radial gradient in $N(\ion{H}{i})/I_{175}$ (see
Fig.~\ref{fig:12}).
{\bf (d)} $I_{1-0}$ as a function of  $I_{175}$ for $R = 35\arcmin
-60\arcmin$.
{\bf (e)} $N_\mathrm{gas}$ as function of $I_{175}$ for $R = 0\arcmin
-35\arcmin$ and
{\bf (f)} for $R = 35\arcmin -60\arcmin$.
Note that the extent of the logarithmic scales varies between panels.
}
\end{figure*}

In Sect.~3.2 we compared the CO and \ion{H}{i} distributions across
the spiral arms. We now compare the general distribution of CO with
that of \ion{H}{i} and each of them with the $175\,\mu$m FIR emission
using the CO and \ion{H}{i} maps smoothed to $90\arcsec$, the resolution
of the FIR map.
At this resolution the CO and \ion{H}{i} arms near $R=10\kpc$ merge to a
broad emission ring, similar to that seen at $175\,\mu$m.
The comparison was restricted to radii $R < 60\arcmin$
(= 13.6~kpc) and intensities above $3\times$ the r.m.s. noise at
$90\arcsec$ resolution. Therefore, the noise in the CO map determined
the selection of points in correlations involving CO intensities and the
noise in the \ion{H}{i} map that in correlations involving intensities
of \ion{H}{i} or total gas.
In spite of this difference, the selected pixels are largely the same;
most pixels are located on the broad emission ring.
To obtain sets of independent data points, i.e.
a beam area overlap of $\le 5\%$, only pixels spaced by
$\ge 1.67\times$ the beamwidth were considered. The resulting
correlations between CO, \ion{H}{i} and the FIR are listed in
Table~\ref{tab:correl} and examples of correlation plots are shown in
Fig.~\ref{fig:15}.

In view of the distinct differences between the CO and \ion{H}{i}
distributions, it is not surprising that their velocity-integrated
intensities are not well correlated (see Fig.~\ref{fig:15}a and b). In
the panel showing the full radial range, $R=0\arcmin - 60\arcmin$, two
maxima occur which correspond to the inner bright CO arm and the main
emission ring, respectively. In the inner arm the molecular fraction is
much larger than in the main ring, and the decrease of this fraction
outwards obviously contributes to the large spread in the points.
Therefore we also analyzed the inner ($R=0\arcmin -35\arcmin$) and outer
($R=35\arcmin - 60\arcmin$) radial ranges separately. In the inner part
$I_{1-0}$ and $I_{\ion{H}{i}}$ are not correlated (the correlation
coefficient $\rho = 0.3$, see Table~\ref{tab:correl}), but for the outer
range some correlation is visible (Fig.~\ref{fig:15}b), with $\rho =
0.61\pm 0.09$. Note that the correlation follows a power law with
exponent $1.7\pm 1.0$. An exponent of 2 is expected if the formation
and destruction rates of H$_2$ are balanced (Reach \&
Boulanger\ \cite{reach+boulanger98}). This occurs in the denser
\ion{H}{i} phase at temperatures near 80~K, which are also found in
M\,31 (Braun \& Walterbos\ \cite{braun+walterbos92}). For a better
check of this possibility in M\,31 this exponent could be determined in
narrow radial ranges at the original resolution of $24\arcsec \times
36\arcsec$ of the \ion{H}{i} map, which may yield better correlations
than obtained here.

The middle panels of Fig.~\ref{fig:15} show that in the interval
$R=35\arcmin - 60\arcmin$ both $I_{\ion{H}{i}}$ and $I_{1-0}$ are well
correlated with the dust emission at $175\,\mu$m. With correlation
coefficients of $\rho = 0.75\pm 0.03$ and $\rho = 0.81\pm 0.02$,
respectively, both correlations are highly significant
(Table~\ref{tab:correl}). The relationship between $I_{\ion{H}{i}}$ and
$I_{175}$ is about linear, whereas that between $I_{1-0}$ and $I_{175}$
follows a power law with exponent $1.6\pm 0.2$ (see
Table~\ref{tab:correl}). This difference reflects the power-law
dependence of  $I_{1-0}$ on $I_{\ion{H}{i}}$ in Fig.~\ref{fig:15}b. The
two branches visible at $\log I_{175} < 1.3$ in Fig.~\ref{fig:15}c are
caused by the strong radial gradient in the apparent
atomic gas--to--dust ratio
$N(\ion{H}{i})/I_{175}$ plotted in Fig.~\ref{fig:13}: points in the
upper branch are at larger radius than points in the lower one. This
branching does not occur in the correlation between $I_{1-0}$ and
$I_{175}$ (Fig.~\ref{fig:15}d), because $2N({\rm H}_2)/I_{175}$
increases much less with $R$ than $N(\ion{H}{i})/I_{175}$.

Correlations between the total gas column density and the $175\,\mu$m
emission for both the inner and the outer part of M\,31 are
better than those between each of the gas components alone and $I_{175}$
(see Table~\ref{tab:correl} and Figs.~\ref{fig:15}e and f). The
correlation for the outer part is very good ($\rho = 0.85\pm 0.02$),
much better than for the inner part ($\rho = 0.68\pm 0.05$). Because in
the inner part $N_\mathrm{gas}$ is dominated by the molecular gas, but
in the outer part by the atomic gas, the power-law exponent in the inner
part ($1.42\pm 0.53$) is larger than in the outer part ($1.12\pm 0.14$).
The latter exponent is consistent with a linear correlation between
$N_\mathrm{gas}$ and $I_{175}$. Linear correlations between gas and
dust in M\,31 were also found by other authors. Pagani et al.
(\cite{pagani+99}) found a linear dependence of the intensity in the
ISOCAM LW2 filter (5.0--$8.0\,\mu$m) on $N_\mathrm{gas}$ in a field
centred on the southern major axis near $X=-30\arcmin$. In a spiral-arm
region centred near $X=-20\arcmin,\ Y=-7\arcmin$ in the SW quadrant,
Neininger et al. (\cite{neininger+98}) obtained a good linear
correlation between $N_\mathrm{gas}$ and the apparent red opacity.
In the Milky Way gas and dust were found to correlate in detail (Bohlin
et al.\ \cite{bohlin+78}; Boulanger \& P\'erault\
\cite{boulanger+perault88}; Boulanger et al.\ \cite{boulanger+96}).

One may ask why the $175\,\mu$m dust emission follows more closely
$N_\mathrm{gas}$ than $N(\ion{H}{i})$ alone, since the gas is mainly
atomic (see Table~\ref{tab:correl}; see also Buat et al.\ \cite{buat+89}).
There are several reasons for this. First, emission from dust in
\ion{H}{i} clouds as well as from dust in H$_2$ clouds contributes to
the $175\,\mu$m emission. Second, according to dust
models (see e.g. Mezger et al.\ \cite{mezger+90}), the dust absorption
coefficient and the dust emissivity per H-atom are two (or more) times
larger in the dense molecular clouds than in the diffuse atomic clouds.
Hence, for equal dust column densities and temperatures, the emission
from dust in CO clouds exceeds that of dust in \ion{H}{i} clouds as
soon as $2N({\rm H}_2)/N(\ion{H}{i}) > 0.3$.

The data presented in Sects. 4.1 and 4.2 show that a radial gradient in
the apparent  gas--to--dust ratio exists that influences the
correlations between the gas components and the dust.
Corrections for \ion{H}{i}-line opacity, better knowledge of the
conversion factor $X_\mathrm{CO}$ and, probably most important, a
determination of the dust temperature as a function of radius are
needed to understand whether this radial gradient reflects a real
increase in the dust--to--gas ratio in M\,31. In the meantime
comparisons of gas and dust on small scales in narrow radial ranges
would be of interest.

%Fig. 15

\subsection{Molecular and total gas mass}

The CO intensity integrated over the area of M\,31 gives an estimate of
the total molecular mass if one assumes a conversion factor between
$I_{1-0}$ and $N({\rm H}_2)$. We integrated the CO intensities in
Fig.~\ref{fig:co-map}a out to a radius of $80\arcmin$ in the plane of
M\,31, which corresponds to $R=18$~kpc. With the same conversion factor
$X_\mathrm{CO}$ as used above we find a molecular mass of $M ({\rm
H}_2) = 3.45\times 10^8\, {\rm M}_{\sun}$. Along the major axis our map
only extends to $R\simeq 55\arcmin$, so we missed the emission near the
major axis between $R = 55\arcmin$ and $75\arcmin$ detected by Koper et
al. (\cite{koper+91}). From Figs.~3b and 10c in Dame et al.
(\cite{dame+93}) we estimate that this contributes about 5\% of the
total. After correcting for these 5\%, we get $M({\rm H}_2) = 3.6\times
10^8\, {\rm M}_{\sun}$ within a radius $R=18$~kpc. This is in excellent
agreement with the value obtained by Dame et al. (\cite{dame+93}), which
is $M({\rm H}_2) = 3.45\times 10^8\, {\rm M}_{\sun}$ after scaling it to
the distance of 780~kpc adopted here. Given that Dame et al. used the
same value for $X_\mathrm{CO}$ as we do, we can also conclude that their
CO intensity scale of radiation temperature, $T_\mathrm{R}$, is in close
agreement with our scale of main beam brightness temperature,
$T_\mathrm{mb}$.

To obtain the atomic gas mass within $R=18$~kpc we integrated the
Westerbork \ion{H}{i} map of Brinks \& Shane (\cite{brinks+shane84})
giving $M (\ion{H}{i}) = 2.6\times 10^9\, {\rm M}_{\sun}$
in the optically thin case.
Thus for $R<18$~kpc the H$_2$ mass is 14\% of the \ion{H}{i} mass and
12\% of the neutral gas mass. The total \ion{H}{i} mass of M\,31 at the
distance of 780~kpc is $4.86\times 10^9\, {\rm M}_{\sun}$ (Cram et al.\
\cite{cram+80}, corrected by Dame et al.\ \cite{dame+93}) which gives a
total neutral gas mass in M\,31 of $M_\mathrm{gas} = 5.2\times 10^9\,
{\rm M}_{\sun}$ and a molecular mass fraction $M ({\rm
H}_2)/M_\mathrm{gas} = 0.07$.

The bulk of the dust in M\,31 is cold dust at a temperature of
$T_\mathrm{d} = 16$~K. Haas et al. (\cite{haas+98}) estimated a total
dust mass of $M_\mathrm{d} = 3.8\times 10^7\, {\rm M}_{\sun}$ at $D =
780$~kpc. However, Schmidtobreick et al. (\cite{schmidto+00}) found
$M_\mathrm{d} = 1.3\times 10^7\, {\rm M}_{\sun}$ using the same data but
a different method of calculation. Hence the apparent  gas--to--dust
mass ratio in M\,31 is $M_\mathrm{gas}/M_\mathrm{d} = 137$--410. If the
\ion{H}{i} masses are corrected for opacity of the \ion{H}{i}, they
increase by 19\% (Braun \& Walterbos\ \cite{braun+walterbos92}) and the
ratio becomes $M_\mathrm{gas}/M_\mathrm{d} = 163$--488. For the Milky
Way Sodroski et al. (\cite{sodroski+94}) obtained a gas--to--dust mass
ratio for the entire Galaxy of $M_\mathrm{gas}/M_\mathrm{d} = 160\pm
60$, similar to the lower estimates for M\,31.

%--------------------------
\section {The CO velocity field}

Figure~\ref{fig:co-map}b shows the first moment of the CO spectral
data cube, i.e. the intensity-weighted mean CO velocity as a function
of position.  This figure can be compared with Fig.~15a of Brinks
\& Burton (\cite{brinks+burton86} -- hereafter B\&B), which shows the
first moment of the \ion{H}{i} data cube. Contrary to \ion{H}{i}, CO
emission becomes very weak at galactocentric radii $R> 18$~kpc,
where the \ion{H}{i} disk starts to flare and to warp (scaled to
the distance of M\,31 of 0.78~Mpc). The velocity field of
Fig.~\ref{fig:co-map}b is therefore little affected by foreground and
background emission from the warp and is more appropriate to derive the
velocity field in the disk than the velocity field of \ion{H}{i}.
Moreover, as M\,31 is at a galactic latitude of $b= -22\degr$,
contamination by foreground Milky Way clouds at velocities of interest
here ($ < -50\kms$) is unlikely. The average deviation between the
\ion{H}{i} and CO mean velocities is only $\rm \simeq 10\kms$, which
may imply that the bulk of the \ion{H}{i} emission in the region
covered by our CO map arises from M\,31's disk, although the different
velocity resolutions ($8.2\kms$ for \ion{H}{i} and $2.6\kms$ for CO) may
contribute to the differences.

The distribution of the mean CO velocities is dominated by the
rotation around the dynamical centre of the galaxy that has a
velocity $V_{\rm sys}= -315\kms$.  Only one small cloud near the centre
of the galaxy and a large cloud complex near the northwestern
minor axis show strong deviations from circular rotation. The small
cloud is located at $X,Y = -0\farcm1,\ 3\farcm1\ (-250\kms )$. Its
non-circular orbit may be caused by the central bar (Berman\
\cite{berman01}; Berman \& Loinard\ \cite{berman+loinard02}).

The cloud complex near the minor axis ($X,Y = 3\arcmin,\ -4\arcmin$),
that appears particularly bright at $175\,\mu$m, shows complex line
profiles that may be related to the complex filamentary structure
visible in Fig.~\ref{fig:7}. Alternatively, they may trace streaming
motions (Muller et al., in prep.).

The width of the observed CO-line profiles varies strongly from position
to position. The average linewidth (second moment of the CO data cube)
computed over the whole disk is $10\pm 5\kms$; it is higher in the arms
($15\pm 5\kms$) and lower at the arm edges and in the interarm regions
($8\pm 4\kms$). The average CO linewidth is in agreement with the
average linewidth of the \ion{H}{i} emission of $8.1\kms$ (at
$1\arcmin$ resolution) arising from the disk (Unwin\ \cite{unwin83}).

Some of the line profiles are very broad (40--65$\kms$, e.g.
Fig.~\ref{fig:spectra}c) and show multiple velocity components. This is
particularly the case for the cloud complexes associated with the dark
clouds D\,47 and D\,39 ($X,Y= -22\farcm48,\ -7\farcm53$ and
$-41\farcm9,\ 8\farcm54$, see Fig.~\ref{fig:spectra}g and i) which show
complex profiles with total widths of 50 and $65\kms$, respectively. We
note that in the case of D\,47 and D\,39 the multiple profiles arise in
the vicinity of bright \ion{H}{ii} regions, but this possible
connection is not a general feature.

Other line profiles appear to be very narrow ($\leq 5\kms$; see
Fig.~\ref{fig:spectra}b). The profiles associated with D\,153 ($X,Y=
-17\farcm8,\ -3\farcm85$, see Fig.~\ref{fig:spectra}f) have
a particularly high peak intensity combined with a narrow
line width of only about $4\kms$.

A detailed analysis of the cloud--to--cloud velocity dispersions inside
the molecular cloud complexes, and of the interplay between density-wave
driven motions and peculiar motions, linked to \ion{H}{ii} regions or
supernova remnants, will be the subject of a forthcoming paper
(Muller et al., in prep.).

%Fig. 16
\begin{figure*}[htb]
\includegraphics[bb = 86 41 484 664,angle=90,width=13cm,clip=]{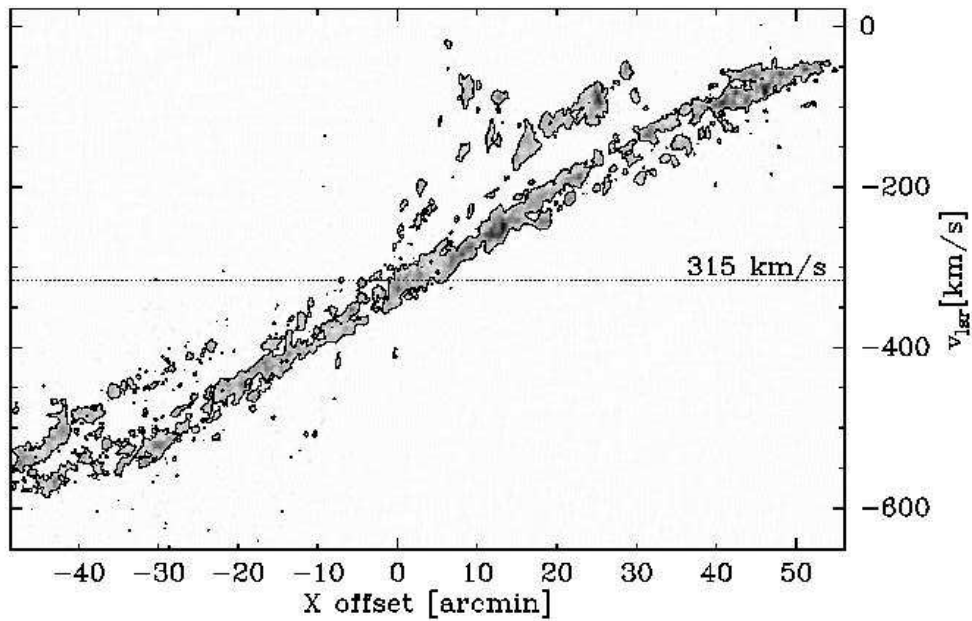}
\caption{\label{fig:16}
Position--velocity diagram along the major axis obtained by adding the
intensities along lines parallel to the $Y$-axis as function of position
on the $X$-axis and velocity. The grey scale represents the intensities
starting at contour level 0.5~K. The dotted line indicates the system
velocity of $-315\kms$.
}
\end{figure*}

The position--velocity diagram of the CO (1--0) intensity integrated
along the minor axis is shown in Fig.~\ref{fig:16}. The dominant
features directly reflect the general velocity field in the disk
(Fig.~\ref{fig:co-map}b) and the fact that most CO emission is
concentrated in spiral arms (Figs.~\ref{fig:co-map}a,\
\ref{fig:spiral}a). Due to the rotation of the M\.31 disk, CO emission
from the arms marked in Fig.~\ref{fig:spiral}a is seen in a broad band
of about $100\kms$ width, stretching from $X \simeq -48\arcmin,\ {\rm v}
\sim -550\kms$ to $X\simeq +55\arcmin,\ {\rm v} = -50\kms$. Most of the
emission in this band is from regions near $R=10$\,kpc. This
large-scale picture agrees of course with that in the low-resolution
survey by Dame et al. (\cite{dame+93}), who also find some emission
near $X = -63,\ {\rm v} = -550\kms$, outside the area of our survey.
The CO spiral arms appear as loops within the range of this band
(compare Braun (\cite{braun91}) for \ion{H}{i}); most clearly seen in
Fig.~\ref{fig:16} is the loop at the most negative $X$ corresponding to
the ``dashed'' spiral arm in Fig.~\ref{fig:spiral}a. Emission from the
inner part of the ``solid'' spiral arm in Fig.~\ref{fig:spiral}a is
clearly visible at $X < 25\arcmin,\ {\rm v} \ga -200\kms$ above the main
band of emission. Extended and relatively strong emission at velocities
around $-80\kms$ and $X$ about $8\arcmin$ to $12\arcmin$ is from clouds
located inside the ``solid'' spiral arm in Fig.~\ref{fig:spiral}a, i.e.
closer to the major axis (compare Fig.~\ref{fig:co-map}b). While we
cannot reliably trace spiral arms from the CO emission in this inner
region, we note that these CO features agree in position and velocity
with the inner loops of the spiral arms identified by Braun (1991) from
\ion{H}{i} surveys. In a similar fashion, most smaller and weaker
features at velocities different from those of the main CO spiral arms
in Figs.~\ref{fig:co-map}b and\ \ref{fig:spiral}a are on other loops of
Braun's \ion{H}{i} spiral arms.

\section{Summary}

The new \element[][12]{CO}(J=1--0)--line survey of the Andromeda Galaxy
presented here covers an area of about $2\degr\times 0\fdg5$, which is
fully sampled with a velocity resolution of $2.6\kms$ and an angular
resolution of $23\arcsec$, the highest angular resolution to date
of a map of this extent.
At the adopted distance of 780\,kpc the linear resolution is $\rm
85\,pc \times 400\,pc$ in the plane of M\,31. The {\em On-the-Fly}
method of observing made it possible to measure nearly 1.7 million
spectra (before gridding) in about 500 hours of effective observing
time on M\,31. The r.m.s. noise in the CO(1--0) line per 1~MHz channel
is 25~mK in the northern half of the map and 33~mK in the southern half.

The velocity-integrated distribution, $I_{1-0}$, and the velocity field
are shown in Fig.~\ref{fig:co-map}. The molecular gas is concentrated in
narrow, filamentary arms between 4 and 12~kpc from the centre with a
maximum near 10~kpc. The inner arm at $R\simeq 5$~kpc is remarkably
bright. Only few clouds, often forming bridge-like structures,
are detected in between the arms above $3\times$ r.m.s. noise of
typically $0.35\Kkms$. The region within 2~kpc from the centre is
almost free of molecular clouds brighter than the sensitivity of this
survey.

The thin CO arms define a two-armed spiral pattern that can be well
described by two logarithmic spirals with constant pitch angle of
$7\degr - 8\degr$. At a resolution of $45\arcsec$ the arm--interarm
contrast reaches a maximum of 20 in $I_{1-0}$ compared to 4 in
\ion{H}{i} for the western bright arms. The \ion{H}{i} arms are much
wider than the molecular arms, and diffuse \ion{H}{i} exists everywhere
in between the arms and at large radii. Few molecular clouds are
visible outside a radius of 16\,kpc.

The velocity field of the molecular gas is very similar to that of the
disk component in \ion{H}{i}. At some positions perturbed velocity
profiles occur that are possibly caused by nearby \ion{H}{ii} regions.

Several selected regions were also observed in the $^{12}$CO(2--1) line.
At a resolution of $23\arcsec$ the line ratios are nearly constant with
mean values of $I_{2-1}/I_{1-0} = 0.5-0.7$ in the arms. These line
ratios are similar to those observed in other galaxies and show no
indication of subthermal excitation.

Averaged radial profiles of the velocity-integrated CO and \ion{H}{i}
distributions show that for a constant conversion factor of
$X_\mathrm{CO} = 1.9\times 10^{20}\, \mbox{mol\, cm}^{-2}\,
(\Kkms )^{-1}$ the molecular fraction of the neutral gas is enhanced
on the spiral arms and decreases radially from about 0.6 on the inner
arms to about 0.3 on the arms at $R\simeq 10$~kpc (see
Fig.~\ref{fig:11}). Along the arms the molecular fraction also varies
considerably. Comparisons with averaged radial profiles of the
$\lambda175\,\mu$m emission, which traces the cold (16\,K) dust,
revealed a strong, continuous increase of the apparent atomic
gas--to--dust ratio of nearly a factor 20 between the centre and
$R\simeq 14$~kpc, whereas the apparent  molecular gas--to--dust ratio
increases by about a factor of 4. The apparent  total gas--to--dust
ratio also increases by about a factor of 20.

The strong apparent gradients in the molecular fraction and the
gas--to--dust ratios influence the cross-correlations between CO,
\ion{H}{i} and FIR($175\,\mu$m) intensities.  In the radial range $R=
35\arcmin - 60\arcmin$ (about 8--14\,kpc) the best correlation exists
between total neutral gas and $175\,\mu$m, followed by
that between CO and $175\,\mu$m and between \ion{H}{i} and $175\,\mu$m.
The relationships between \ion{H}{i} and $175\,\mu$m and total gas and
$175\,\mu$m are close to linear, whereas that between CO and $175\,\mu$m
is a power law with exponent 1.6 due to a possible non-linearity between
CO and \ion{H}{i}. In the inner part of M\,31, $R < 35\arcmin$ (but
outside the nuclear area), only total gas and FIR($175\,\mu$m) are
reasonably well correlated (see Table~\ref{tab:correl}).

The total molecular mass of M\,31 within a radius of 18~kpc is $\rm
3.6\times 10^8\, M_{\sun}$, using the above-mentioned value of
$X_\mathrm{CO}$. As the total \ion{H}{i} mass (without correction for
opacity) is $\rm 4.86\times 10^9\, M_{\sun}$, the total mass of the
neutral gas is $\rm 5.2\times 10^9\, M_{\sun}$. The total mass of the
cold dust is $\rm (1.3-3.8)\times 10^9\, M_{\sun}$, hence the total
gas--to--dust mass ratio is 410--137. The lower value is in agreement
with the Galactic one.

The wealth of information contained in our new $^{12}$CO(1--0) survey of
M\,31 enables a number of new investigations into the physical
relationships between molecular gas, atomic hydrogen gas, cold and warm
dust, ionized gas, relativistic electrons and magnetic fields. Such
studies will be the subject of forthcoming papers.

The data shown in Fig.~\ref{fig:co-map} can be obtained from M.~Gu\'elin
({\tt guelin@iram.fr}).

\begin{acknowledgements}
C.N. acknowledges support from the Studienstiftung des Deutschen
Volkes. We thank W. Brunswig, C. Kramer, G. Paubert, J. Schraml and A.
Sievers for their important contributions to the {\em On-the-Fly}
observing mode at the IRAM 30-m telescope and P. Hoernes for
observations and software adjustment. S. Muller kindly communicated to
us some results of his PhD thesis prior to publication. We thank M.-C.
Willaime for sending us an updated version of the ISOCAM LW6 map that we
used in Fig.~\ref{fig:7}. The extensive comments and suggestions
of an anonymous referee were a great help in improving the manuscript.
This work was started when M.G. was visiting the MPIfR as a von
Humboldt Fellow.
\end{acknowledgements}

\end{document}